\documentclass[useAMS,usenatbib]{mn2e}

\usepackage[T1]{fontenc}
\usepackage[latin9]{inputenc}
\usepackage{color}
\usepackage{float}
\usepackage{url}
\usepackage{amsmath}
\usepackage{amssymb}
\usepackage{graphicx}
\usepackage{setspace}
\usepackage{esint}

\title[Sparsely Sampling the Sky: A Bayesian Experimental Design Approach]{Sparsely Sampling the Sky: A Bayesian Experimental Design Approach}
\author[P. Paykari and A. H. Jaffe]{P. Paykari$^{1}$\thanks{E-mail:
paniez.paykari@cea.fr; a.jaffe@ic.ac.uk} and A. H.
Jaffe$^{2}$\\
$^{1}$Laboratoire AIM, UMR CEA-CNRS-Paris 7, Irfu, SAp/SEDI, Service d'Astrophysique, CEA Saclay, F-91191 GIF-
SUR-YVETTE CEDEX, France.\\
$^{2}$Department of Physics, Blackett Laboratory, Imperial College, London SW7 2AZ, United Kingdom}

\begin{document}
\date{}
\pagerange{\pageref{firstpage}--\pageref{lastpage}} \pubyear{2002}
\maketitle
\label{firstpage}

\begin{abstract}
The next generation of galaxy surveys will observe millions of galaxies
over large volumes of the universe. These surveys are expensive both
in time and cost, raising questions regarding the optimal investment
of this time and money. In this work we investigate criteria for selecting
amongst observing strategies for constraining the galaxy power spectrum
and a set of cosmological parameters. Depending on the parameters of
interest, it may be more efficient to observe a larger, but sparsely
sampled, area of sky instead of a smaller contiguous area. In this
work, by making use of the principles of Bayesian Experimental Design,
we will investigate the advantages and disadvantages of the sparse
sampling of the sky and discuss the circumstances in which a sparse
survey is indeed the most efficient strategy. For the Dark Energy
Survey (DES), we find that by sparsely observing the same area in
a smaller amount of time, we only increase the errors on the parameters
by a maximum of 0.45\%. Conversely, investing the same amount of time
as the original DES to observe a sparser but larger area of sky we
can in fact constrain the parameters with errors reduced by 28\%. 
\end{abstract}

\begin{keywords}
cosmology
\end{keywords}

\section{Introduction}

The measurements of the cosmological parameters heavily
rely on accurate measurements of power spectra. Power spectra describe
the spatial distribution of an isotropic random field, defined as
the Fourier transform of the spatial correlation function. The perturbations
in the universe can be described statistically using the correlation
function $\xi(r)$ between two points, which depends only
on their separation $r$ (when isotropy is assumed)\footnote{Note that we use underlined symbols to denote vectors and bold symbols for matrices.}; 
\begin{equation}
\xi(r)\equiv\left\langle \delta(\underbar{x})\delta(\underbar{x}+\underline{r})\right\rangle \;,
\end{equation}
where $\delta(\underbar{x})=\left(\rho(\underbar{x})-\bar{\rho}\right)/\bar{\rho}$
measures the continuous over-density, where $\rho(\underline{x})$ is the
density at position $\underline{x}$ and $\bar{\rho}$ is the average density. 
The power spectrum $P(k)$, which is the Fourier transform of the
correlation function, is enough to define the perturbations completely
when the perturbations are assumed uncorrelated Gaussian random fields
in the Fourier space. Power spectra (or correlation functions) are
what the surveys actually measure, from which cosmological parameters are inferred. 
These spectra are normally a convolution
of the primordial power spectrum (which measures the statistical distribution
of perturbations in the early universe) and a transfer function which
depends on the cosmological parameters. Hence accurate measurements
of the power spectra from surveys are very important for accurate measurements of the cosmological
parameters. 

The most important observed spatial power spectrum for
cosmology is the galaxy power spectrum; the Fourier transform of the galaxy correlation function, which was first formulated by \citet{Peebles1973}. A galaxy survey lists the measured positions of the observed galaxies.  As proposed by Peebles, these positions are modelled as a random Poissonian point source, where the galaxy density is modulated by the fluctuations in the underlying matter distribution and the selection effects. The selection function of the survey is described by $\bar{n}(\underbar{x})$, which is the expected galaxy density at position $\underbar{x}$ in the absence of clustering. The fluctuations in the underlying matter density are given by 
$\delta(\underbar{x})$, as described previously. The the galaxy number over-density $n(\underbar{x})$, which is the observed quantity, is related to the matter over-density via the bias $b$ \citep{kaiser1984_bias} --- galaxies trace dark matter up to this $b$
factor. We define the galaxy power spectrum $P_{g}(k)$ as 
\begin{equation}
P_{g}(k)=2\pi^{2}\cdot b^{2}(k)\cdot k\cdot T^{2}(k)\cdot P_{p}(k)\;,
\end{equation}
where $P_{p}(k)$ is the primordial power spectrum $P_{p}(k)=A_{s}k^{n_{s}-1}$.
The transfer function $T(k)$ further depends upon the cosmological
parameters (e.g., the matter density $\Omega_m$, the scalar spectral index, $n_s$, etc.) responsible for the evolution of the
universe. The bias $b$ relates the galaxy power spectrum
to the matter power spectrum, as explained above.

This power spectrum is very rich in
terms of constraining a large range of cosmological parameters. On
large scales this spectrum probes structure which is less affected
by clustering and evolution. Hence these scales are still in the linear
regime and have a ``memory'' of the initial state. The information
from these regimes are, therefore, the cleanest since the Big Bang and any
knowledge on these large scales would shed light on the physics of
early universe and hence the primordial power spectrum. On intermediate scales
the spectrum provides us with information about the evolution of the
universe since the Big Bang; for example the matter-radiation equality
which is responsible for the peak of the galaxy spectrum. The matter-radiation
equality is a unique point in the history of the evolution, giving
information about the amount of matter and radiation in the universe.
On relatively small scales there is a great deal of information about
galaxy clustering via the Baryonic Acoustic Oscillations (BAO) which
encode a characteristic scale; the sound horizon at the time of recombination.
Therefore, measuring the galaxy power spectrum on a large range of
scales can help us constrain the cosmological parameters responsible
for the evolution of the universe as well as the ones of its initial
state.

Accurate measurements of the galaxy power spectrum depend on two main
factors; the Poisson noise and the cosmic variance. To overcome the
Poisson noise, surveys aim to maximise the number of galaxies observed.
The impressive constraints on cosmological parameters from previous
and current surveys, such as the 2dF \citep{2dF} and SDSS \citep{sdss},
has motivated even more ambitious future surveys such as DES \citep{DES}
and Euclid \citep{Euclid}, aiming to observe millions of galaxies
over large volumes of the universe. Considering the large investments
in time and money for these surveys, one wants to ask what is really
the optimal survey strategy! In this work we want to investigate this
exact questions and find the optimal strategy for galaxy surveys such
as DES and Euclid. 

In this era of cosmology where the statistical errors have reduced
greatly and are now comparable with systematics, observing, for example,
a greater number of galaxies may not necessarily improve our results.
We need to devise more strategic ways to make our observations and
take control of our systematics. For example, to investigate larger
scales, it may be more efficient to observe a larger, but sparsely
sampled, area of sky instead of a smaller contiguous area. In this
case we would gather a larger density of states in Fourier space,
but at the expense of an increased correlation between different scales
--- aliasing. This would smooth out features on these scales and decrease
its significance if any observed. Here, by making use of Bayesian
Experimental Design we will investigate the advantages and disadvantages
of the sparse sampling and verify if a complete contiguous survey
is indeed the most efficient way of observing the sky for our purposes.
The parameter of interest here is the galaxy power spectrum itself
and a set of cosmological parameters that depend on this spectrum.

Some previous work on sparse sampling includes \citet{Kaiser-Sparse} and \citet{Blakeetal06};
\citet{Kaiser-Sparse} shows that measuring the large scale
correlation function from a complete magnitude-limited redshift survey
is actually not the most efficient approach. Instead, sampling a fraction
of galaxies randomly, but to a fainter magnitude limit, will improve
the constraints of the correlation function measurements significantly,
for the same amount of observing time. \citet{Blakeetal06}
have shown that a sparse-sampling (achieved by a non-contiguous telescope
pointings or, for a wide-field multi-object spectrograph,
by having the fibres distributed randomly across the field-of-view)
is preferred when the angular size of the sparse observed patches
is much smaller than angular scale of the features in the power spectrum
(the acoustic features).

\section{Bayesian Experimental Design and Figure-of-Merit}

Bayesian methods have recently been used in cosmology for model comparison
and for deriving posterior probability distributions for parameters
of different models. However, Bayesian statistics can do even more
by handling questions about the performance of future experiments,
based on our current knowledge \citep{Liddleetal06,Trotta07,Trotta07-BayesFactor}.
For example, \citet{PBKBNG-BayesExp} use
a Bayesian approach to constrain the dark energy parameters by optimising the
Baryon Acoustic Oscillations (BAO) surveys. By searching through a
survey parameter space (which includes parameters such as redshift
range, number of redshift bins, survey area, observing
time, etc.) they find the optimal survey with respect to the dark energy
equation-of-state parameters. Here we will use this strength of Bayesian
statistics for optimising the strategy to observe the sky for galaxy
surveys. There are three requirements for such an
optimisation; 1. specify the parameters that define the experiment which need to be optimised for an optimal survey; 2. specify the parameters
to constrain, with respect to which the survey is optimised; 3. specify
a quantity of interest, generally called the figure of merit (FoM),
associated with the proposed experiment. The choice of the FoM depends
on the questions being asked, as will be explained later in the text. We then want to extrimise the FoM subject to constraints imposed
by the experiment or by our knowledge about the nature of the universe.
Below, we will explain the procedure.

Assume $e$ denotes the different experimental designs that we can
implement and $M^{i}$ are the different models under consideration with
their parameters $\theta^{i}$. Assume that experiment $o$ has been
performed, so that this experiment's posterior $P(\theta|o)$ forms
our prior probability function for the new experiment. The FoM will depend on the set of parameters under investigation, the
performed experiment (data) and the characteristics of the future
experiment; $U(\theta,e,o)$. From the utility we can build
the expected utility $E\left[U\right]$ as
\begin{equation}
E[U|e,o]=\sum_{i}P(M^{i}|o)\int d\hat{\theta}^{i}\; U(\hat{\theta}^{i},e,o)P(\hat{\theta}^{i}|o,M^{i})\:,
\end{equation}
where $\hat{\theta}^{i}$ represent the fiducial parameters for model
$M^{i}$. This says: If a set of fiducial parameters, $\hat{\theta}$,
correctly describe the universe and we perform an experiment $e$,
then we can compute the utility function for that experiment, $U(\hat{\theta},e,o)$.
However, our knowledge of the universe is described by the current
posterior distribution $P(\hat{\theta}|o)$. Averaging the utility
over the posterior accounts for the present uncertainty in the parameters
and summing over all the available models would account for the uncertainty
in the underlying true model.
The aim is to select an experiment that extremises the utility function (or its expectation). 
The utility function takes into account the current
models and the uncertainties in their parameters and, therefore, extremising
it takes into account the lack of knowledge of the true model of the
universe.

One of the common choices for the FoM is some form of function
of the Fisher matrix, which is the expectation of the inverse covariance
of the parameters in the Gaussian limits (We will explain in the next
section how a Fisher matrix is obtained in more detail.). One can refer to the Dark Energy Task Force (DETF) FoM, that use Fisher-matrix techniques to investigate how 
well each model experiment would be able to restrict the dark energy parameters $w_0$, $w_a$, $\Omega_{DE}$ for their purposes. Three common FoMs, which we will be using as well, are
\begin{itemize}
\item A-optimality $=\log(\textrm{trace}(\mathbf{F}))$\\
trace of the Fisher matrix (or its $\log$) and is proportional to
sum of the variances. This prefers a spherical error region, but may
not necessarily select the smallest volume. 
\item D-optimality $=\log\left(\left|\mathbf{F}\right|\right)$ \\
determinant of the Fisher matrix (or its $\log$), which measures
the inverse of the square of the parameter volume enclosed by the
posterior. This is a good indicator of the overall size of the error
over all parameter space, but is not sensitive to any degeneracies
amongst the parameters. 
\item Entropy (also called the Kullback-Leibler divergence) 
\begin{eqnarray}
E & = & \int d\theta\; P(\theta|\hat{\theta},e,o)\log\frac{P(\theta|\hat{\theta},e,o)}{P(\theta|o)}\nonumber \\
 & = & \frac{1}{2}\left[\log\left|\mathbf{F}\right|-\log|\mathbf{\Pi}|-\textrm{trace}(\mathbb{I}-\mathbf{\Pi}\mathbf{F}^{-1})\right]\,,
\end{eqnarray}
where $P(\theta|\hat{\theta},e,o)$ is the posterior distribution
with Fisher matrix $\mathbf{F}$ and $P(\theta|o)$ is the prior distribution
with Fisher matrix $\mathbf{\Pi}$. The entropy forms a nice compromise
between the A-optimality and D-optimality. Note that these are the utility functions, not the `expected' utility functions. In our current models of the universe, we do not expect a significant difference between the parameters of the same model. However, this will be investigated in a future work, where we will explicitly use expected utility functions. In the next section we
will explain how a Fisher matrix is formulated.
\end{itemize}

\section{Fisher Matrix Analysis}

The Fisher matrix is generally used to determine the sensitivity of
a particular survey to a set of parameters and has been largely used
for optimisation (and forecasting). Consider the likelihood function
for a future experiment with experimental parameters $e$, $\mathcal{L}(\theta|e)\equiv P(D_{\hat{\theta}}|\theta,e)$,
where $D_{\hat{\theta}}$ are simulated data from the future experiment
assuming that $\hat{\theta}$ are the true parameters in the given
model. We Taylor expand the log-likelihood around its maximum value:
\begin{equation}
\ln\mathcal{L}(\theta|e)=\ln\mathcal{L}(\theta^{ML})+\frac{1}{2}\sum_{ij}(\theta_{i}-\theta_{i}^{ML})\frac{\partial^{2}\ln\mathcal{L}}{\partial\theta_{i}\partial\theta_{j}}(\theta_{j}-\theta_{j}^{ML})\:,
\end{equation}
where the first term is a constant and only affects the height of
the function, the second term describes how fast the likelihood function
falls around the maximum. The Fisher matrix is
defined as the ensemble average of the \textit{curvature} of the likelihood
function $\mathcal{L}$ (i.e., it is the average of the curvature
over many realisations of signal and noise);
\begin{eqnarray}
F_{ij} & = & \left\langle \mathcal{F}\right\rangle =\left\langle -\frac{\partial^{2}\ln\mathcal{L}}{\partial\theta_{i}\partial\theta_{j}}\right\rangle \label{eq:General_FM}\\
 & = & \frac{1}{2}\textrm{trace}[C_{,i}C^{-1}C_{,j}C^{-1}]\:,
\end{eqnarray}
where the second line is appropriate for a Gaussian
distribution with correlation matrix $C$ determined by the parameters
$\theta_{i}$, and $\mathcal{L}$ is the likelihood function. The
inverse of the Fisher matrix is an approximation of the covariance
matrix of the parameters, by analogy with a Gaussian distribution
in the $\theta_{i}$, for which this would be exact. The Cramer-Rao
inequality%
\footnote{It should be noted that the Cramer-Rao inequality is a statement about
the so-called ``Frequentist'' confidence intervals and is not strictly
applicable to ``Bayesian'' errors. %
} states that the smallest \textcolor{black}{frequentist} error measured,
for $\theta_{i}$, by any unbiased estimator (such as the maximum
likelihood) is $1/\sqrt{F_{ii}}$ and $\sqrt{(F^{-1})_{ii}}$, for
non-marginalised and marginalised%
\footnote{Integration of the joint probability over other parameters.%
} one-sigma errors respectively. The derivatives in Equation \ref{eq:General_FM}
generally depend on where in the parameter space they are calculated
and hence it is clear that the Fisher matrix is function of the fiducial
parameters.

The Fisher matrix allows us to estimate the errors on parameters without
having to cover the whole parameter space (but
of course will only be appropriate so long as the derivatives are
roughly constant throughout the space). So, a Fisher matrix analysis is equivalent to the assumption of a Gaussian distribution about the peak of the likelihood \citep[e.g.][]{bjk}.
It also makes the calculations easier. For example,
if we are only interested in a subset of parameters, then marginalising
over unwanted parameters is just the same as inverting the Fisher
matrix, taking only the rows and columns of the wanted parameters
and inverting the smaller matrix back. It is also very straightforward
to combine constraints from different independent parameters: we just
sum over the Fisher matrices of the experiments (remember Fisher matrix
is the $\log$ of the likelihood function).

We further note, as in all uses of the Fisher matrix, that any results
thus obtained must be taken with the caveat that these relations only
map onto realistic error bars in the case of a Gaussian distribution,
usually most appropriate in the limit of high signal-to-noise ratio
and/or relatively small scales, so that the conditions of the central
limit theorem obtain. As long as we do not find extremely degenerate
parameter directions, we expect that our results will certainly be
indicative of a full analysis, using simulations and techniques such
as Bayesian Experimental Design \citep{T-BayesExp}.

\subsection{Fisher Matrix for Galaxy Surveys}

We follow the approach of \citet{tegmark1997} to define the pixelisation
for galaxy surveys. First we define the data in pixel $i$ as 
\begin{equation}
\Delta_{i}\equiv\int d^{3}x\psi_{i}\left(\underbar{x}\right)\left[\frac{n\left(\underbar{x}\right)-\bar{n}}{\bar{n}}\right]\,,
\end{equation}
where $n(\underline{x})$ is the galaxy density at position $\underline{x}$
and $\bar{n}$ is the expected number of galaxies at that position.
The weighting function, $\psi_{i}(\underline{x})$, which determines
the pixelisation (and is sensitive to the shape of the survey as you
will see later), is defined as a set of Fourier pixels
\begin{eqnarray}
\psi_{i}(\underline{x})=\frac{e^{\iota\underline{k}_{i}.\underline{x}}}{V}\times\begin{cases}
1 & \,\vec{x}\,\,\textnormal{ inside\,\,\ survey\,\,\ volume}\\
0 & \,\textnormal{otherwise}
\end{cases}\,,\label{eq:weighting_fn}
\end{eqnarray}
where $V$ is the volume of the survey. Here we have divided the volume
into sub-volumes, each being much smaller than the total volume of
the survey, but being large enough to contain many galaxies. This
means $\Delta_{i}$ is the fractional over-density in pixel $i$.
Using this pixelisation we can define a covariance matrix as
\begin{equation}
\left\langle \Delta_{i}\Delta_{j}^{*}\right\rangle =C=(C_{S})_{ij}+(C_{N})_{ij}\,,
\end{equation}
where $C_{S}$ and $C_{N}$ are the signal and noise covariance matrices
respectively and are assumed independent of each other. The signal
covariance matrix can be defined as 
\begin{eqnarray}
(C_{S})_{ij} & = & \left\langle \Delta_{i}\Delta_{j}^{*}\right\rangle \nonumber \\
& = & \int d^{3}xd^{3}x^{\prime}\;\psi_{i}(\underbar{x})\psi_{j}^{*}(\underbar{x}^{\prime}) \nonumber \\
& \; & \left\langle \frac{n(\underbar{x})-\bar{n}}{\bar{n}}\cdot\frac{n(\underbar{x}^{\prime})-\bar{n}}{\bar{n}}\right\rangle \,.\label{eq:C_S_ij1}
\end{eqnarray}
By equating the number over-density $\left(n(\underbar{x})-\bar{n}\right)/\bar{n}$ to the continuous over-density $\delta(\underbar{x})=\left(\rho(\underbar{x})-\bar{\rho}\right)/\bar{\rho}$ we obtain
\begin{eqnarray}
(C_{S})_{ij} & = & \int\frac{d^{3}k}{(2\pi)^{3}}P(k)\tilde{\psi}_{i}(\underline{k})\tilde{\psi}_{j}^{*}(\underline{k})\nonumber \\
 & = & \int\frac{dk}{(2\pi)^{3}}k^{2}P(k)\int d\Omega_{k}\;\tilde{\psi}_{i}(\underline{k})\tilde{\psi}_{j}^{*}(\underline{k})\nonumber \\
 & = & \int\frac{dk}{(2\pi)^{3}}k^{2}P(k)W_{ij}(k)\,,\label{eq:C_S_ij2}
\end{eqnarray}
where $\tilde{\psi}_{i}(\underline{k})$ is the Fourier transform
of $\psi_{i}(\underline{x})$ and the window function $W_{ij}(k)$
is defined as the angular average of the square of the Fourier transform
of the weighting function. With the same approach, the noise covariance
matrix --- which is due to Poisson shot noise --- is given by
\begin{eqnarray}
(C_{N})_{ij} & = & \left\langle N_{i}N_{j}^{*}\right\rangle _{\textnormal{Noise}}\nonumber \\
 & = & \int d^{3}xd^{3}x^{\prime}\psi_{i}\left(\underbar{x}\right)\psi_{j}^{*}\left(\underbar{x}^{\prime}\right)\frac{1}{\overline{n}}\delta_{D}\left(\underbar{x}-\underbar{x}^{\prime}\right)\nonumber \\
 & = & \int\frac{d^{3}k}{(2\pi)^{3}}\frac{1}{\overline{n}}\tilde{\psi}_{i}(\underline{k})\tilde{\psi}_{j}^{*}(\underline{k})\nonumber \\
 & = & \int\frac{dk}{(2\pi)^{3}}k^{2}\frac{1}{\overline{n}}\int d\Omega_{k}\tilde{\psi}_{i}(\underline{k})\tilde{\psi}_{j}^{*}(\underline{k})\nonumber \\
 & = & \frac{1}{\overline{n}}\int\frac{dk}{(2\pi)^{3}}k^{2}W_{ij}(k)\ .
\end{eqnarray}
The design of the survey will shape the form of the weighting function
in Equation \ref{eq:weighting_fn}, which will be discussed in the
next section. 

This prescription gives us a data covariance matrix for a galaxy survey.
What we actually need is a Fisher matrix for the parameters we are
interested in. For this we will use Equation \ref{eq:General_FM}
above, which defines the Fisher matrix of parameters in terms of the
inverse of the data covariance matrix and its differentiation with
respect to the parameters of interest. We are interested in the galaxy
power spectrum and hence the differentiation of the covariance matrix
in Equation \ref{eq:General_FM} is taken with respect to the bins
of this power spectrum. As the noise covariance matrix does not depend
on the power spectrum, we only need to differentiate the signal covariance
matrix in Equation \ref{eq:C_S_ij2}. Taking the galaxy power spectrum
as a series of top-hat bins 
\begin{equation}
P(k)=\sum_{B}w_{B}(k)P_{B}\quad\begin{cases}
w_{B}=1 & \,\, k\in B\\
0 & \,\textnormal{otherwise}
\end{cases}\,,
\end{equation}
where $P_{B}$ is the power in each bin, the differentiation takes
the form
\begin{equation}
\frac{\partial(C_{S})_{ij}}{\partial P(k)}=\int_{k_{B}^{min}}^{k_{B}^{max}}\frac{dk}{(2\pi)^{3}}k^{2}W_{ij}(k)\,.
\end{equation}

We insert this and the inverse of the data covariance matrix into
Equation \ref{eq:General_FM} to get a Fisher matrix for the galaxy
power spectrum bins. To get a Fisher matrix for the cosmological parameters
one can use the parameters Jacobian
\begin{equation}
F_{\alpha\beta}=\sum_{ab}F_{ab}\frac{\partial P_{a}}{\partial\lambda_{\alpha}}\frac{\partial P_{b}}{\partial\lambda_{\beta}}\;.
\end{equation}
where $F_{ab}$ is the galaxy spectrum Fisher matrix and $F_{\alpha\beta}$
is the Fisher matrix for the cosmological parameters $\lambda_{\alpha}$
and $\lambda_{\beta}$.

\section{Survey Design}

We will investigate the FoM of a sparse design to that of a contiguous
survey, which we have chosen to b\textcolor{black}{e similar to that
of t}he Dark Energy Survey (DES).

\subsection{Dark Energy Survey (DES) }

The Dark Energy Survey (DES)%
\footnote{\url{http://www.darkenergysurvey.org/}%
} \citep{DES} is designed to probe the origin of the accelerating universe
and help uncover the nature of dark energy. Its digital camera, DECam,
is mounted on the Blanco 4-meter telescope at Cerro Tololo Inter-American
Observatory in the Chilean Andes. Starting in December 2012 and continuing
for five years, DES will catalogue 300 million galaxies in the southern
sky over an area of 5000 square degrees and a redshift range of $0.2<z<1.3$.
In the next section we will explain how we `sparsify' the DES survey
for our purposes. 

Here, we use a flat-sky approximation. Euclid, with a survey area of  $20,000$ square degrees should be treated on the full sky and is not investigated here. Nonetheless we expect qualitatively similar results to DES.

\subsection{Sparse Design }

\begin{figure}
\includegraphics[width=\columnwidth]{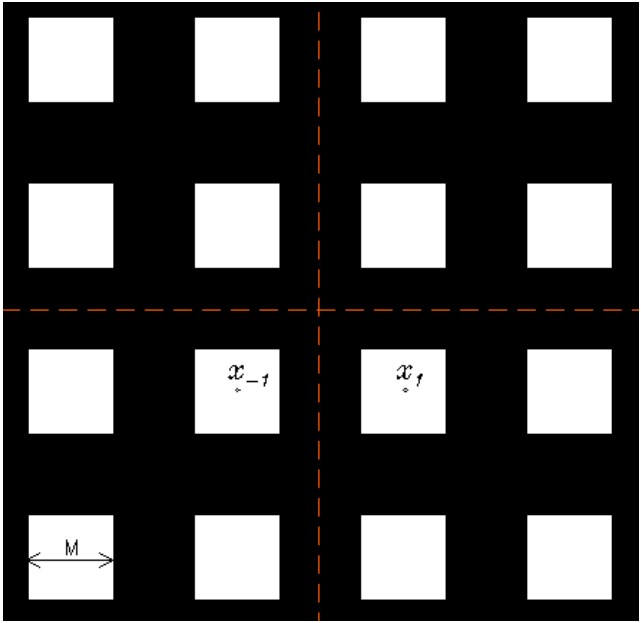}
\caption{Design of the mask on the sky to sparsely sample the sky. A regular
grid with $n$ patches of size $M$ (note that we are observing through these patches --- white squares in the Figure), placed at constant distances from
one another at $x_{i}$ and $y_{j}$. The total \textit{observed} area is the sum of the areas of all the patches, $n \times M^2$, and total \textit{sampled} area is the total area which bounds both the masked and the unmasked areas, $V$. Hence the fraction of sky observed is $f=(n\times M^2)/A_{\textrm{tot}}$. Also, note that we are assuming a flat-sky approximation.\label{fig:Design}}
\end{figure}

For simplicity, we will design the sparsely sampled area of the sky
as a regular grid of $n_{p}\times n_{p}$ square patches of size $M\times M$
--- Figure \ref{fig:Design}. We therefore define the structure on
the sky as a top-hat in both $x$ and $y$ directions 
\begin{eqnarray}
\sum_{n}\Pi(x-x_{n}) & = & \begin{cases}
1 & \,0<|x-x_{n}|<M/2\\
0 & \,\textnormal{otherwise}
\end{cases}\,,\\
\sum_{m}\Pi(y-y_{m}) & = & \begin{cases}
1 & \,0<|y-y_{m}|<M/2\\
0 & \,\textnormal{otherwise}
\end{cases}\,,
\end{eqnarray}
where $x_{i}$ and $y_{j}$ mark the centres of the patches in our
coordinate system. In the $z$ direction we use the step function,
which is defined as:
\begin{eqnarray}
\Theta(z) & = & \begin{cases}
1 & z > 0 \\
0 & \,\textnormal{otherwise}
\end{cases}\,.
\end{eqnarray}
With this design the weight function in equation \ref{eq:weighting_fn}
takes the form: 
\begin{eqnarray}
\tilde{\psi}_{i}(\underline{k}) & = & \int d^{3}x\; e^{\imath(\underline{k}_{i}-\underline{k}).\underline{x}}  \;\;\;\times\; \nonumber \\
 & \; & \;\;\;\;\sum_{n}\Pi(x-x_{n})\sum_{m}\Pi(y-y_{m}) \;\;\;\times\; \nonumber \\
 & \; & \;\;\;\;\Theta\left(z+\frac{L}{2}\right)\Theta\left(\frac{L}{2}-z\right) \;\times\frac{1}{V} \nonumber \\
 & = & \int dxe^{\imath q_{x}x}\sum_{n}\Pi(x-x_{n}) \;\;\;\times\; \nonumber \\
 & \; & \int dye^{\imath q_{y}y}\sum_{m}\Pi(y-y_{m}) \;\;\;\times\; \nonumber \\
 & \; & \int dze^{\imath q_{z}z}\Theta\left(z+\frac{L}{2}\right)\Theta\left(\frac{L}{2}-z\right) \;\times\frac{1}{V} \nonumber \\
 & = & \text{sinc}\left(q_{x}\frac{M}{2}\right)\sum_{n}2\cos\left(q_{x}x_{n}\right) \;\;\;\times\; \nonumber \\
 & \; & \text{sinc}\left(q_{y}\frac{M}{2}\right)\sum_{m}2\cos\left(q_{y}y_{m}\right) \;\;\;\times\; \nonumber \\
 & \; & \text{sinc}\left(q_{z}\frac{L}{2}\right) \;\times\frac{M^{2}L}{V} \ ,
\label{eq:weighting_fn2}
\end{eqnarray}
where {$\underline{q}=\underline{k}_{i}-\underline{k}$},
$q_{x}=q\sin\theta\cos\phi$, $q_{y}=q\sin\theta\sin\phi$, $q_{z}=q\cos\phi$
and $d\mu=d\cos\theta$. The volume $V$ is the \textit{total} sparsely
sampled volume, $M$ is the size of the observed patch on the surface
of the sky and $L$ is the observed depth. The last equality in the
above equation uses the \textit{Dirichlet Kernel}
\begin{equation}
D_{n}(x)=\sum_{k=-n}^{n}e^{ikx}=1+2\sum_{k=1}^{n}\cos(kx)\,,
\end{equation}
which can be used due to the symmetry of the design. The window
function, defined in Equation \ref{eq:C_S_ij2}, now takes the form

\begin{eqnarray}
W_{ij}(k) & = & \int_{-1}^{1}\frac{d\mu}{2}\int_{0}^{2\pi}\frac{d\phi}{2\pi}\tilde{\psi}(\underline{k}_{i}-\underline{k})\tilde{\psi^{*}}(\underline{k}_{j}-\underline{k})\nonumber \\
 & = & \int_{-1}^{1}\frac{d\mu}{2}\int_{0}^{2\pi}\frac{d\phi}{2\pi}\times\left(\frac{M^{2}L}{V}\right)^{2}\times\nonumber \\
 &  & \;\;\;\;\;\;\;\text{sinc}\left(q_{x}\frac{M}{2}\right)\sum_{n}2\cos(q_{x}x_{n}) \times\nonumber \\
 &  & \;\;\;\;\;\;\;\text{sinc}\left(q_{x}^{\prime}\frac{M}{2}\right)\sum_{n^{\prime}}2\cos(q_{x}^{\prime}x_{n^{\prime}})\times\nonumber \\
 &  & \;\;\;\;\;\;\;\text{sinc}\left(q_{y}\frac{M}{2}\right)\sum_{m}2\cos(qy_{m})\times\nonumber \\
 &  & \;\;\;\;\;\;\;\text{sinc}\left(q_{y}^{\prime}\frac{M}{2}\right)\sum_{m^{\prime}}2\cos(q_{y}^{\prime}y_{m^{\prime}})\times\nonumber \\
 &  & \;\;\;\;\;\;\;\text{sinc}\left(q_{z}\frac{L}{2}\right)\text{sinc}\left(q_{z}^{\prime}\frac{L}{2}\right)\;.\label{eq:Window Function}
\end{eqnarray}
Note that there are two scales that control the behaviour of the window
function; one is the size of the patches, $M$, and the other is their
distance from one another, $x_{i}$. We will investigate the influence
of both of these scales on the FoM by trying two different configurations,
discussed in the next section.
In case of the contiguous sampling of the sky where we are observing
through a contiguous square, the window function takes the form of
one single big patch, as shown below 
\begin{eqnarray}
W_{ij}(k) & = & \int_{-1}^{1}\frac{d\mu}{2}\int_{0}^{2\pi}\frac{d\phi}{2\pi}\times\nonumber \\
 &  & \;\;\;\;\;\;\;\text{sinc}\left(q_{x}\frac{M}{2}\right)\text{sinc}\left(q_{x}^{\prime}\frac{M}{2}\right)\times\nonumber \\
 &  & \;\;\;\;\;\;\;\text{sinc}\left(q_{y}\frac{M}{2}\right)\text{sinc}\left(q_{y}^{\prime}\frac{M}{2}\right)\times\nonumber \\
 &  & \;\;\;\;\;\;\;\text{sinc}\left(q_{z}\frac{L}{2}\right)\text{sinc}\left(q_{z}^{\prime}\frac{L}{2}\right)\;.
\end{eqnarray}
which is a square cylinder.

\subsection{Sparsifying DES}

We divide the total area of DES into small square patches, as explained
in the design of the mask previously. There are two ways to sparsify
this area; 
\begin{itemize}
\item Constant Total Area (full sampled area stays constant)\\
In this setting we keep the patches at a constant position and gradually
decrease their size. Therefore, the total \textit{sampled}
\footnote{This is the total area including both the masked and unmasked areas.} area is kept constant, while the total \textit{observed} area decreases as
the patch sizes decrease. The patches are placed at $60\textnormal{Mpc}$
from one another; this scale is about half of the
scale of the BAO Scales, which is $\sim120\textnormal{Mpc}$. The
patches are placed at half this scale to capture the BAO features
at best. This restricts the maximum size of the patches
to be $60\textnormal{Mpc}$ for $f=1$. We then shrink them
from $60\textnormal{Mpc}$ to $10\textnormal{Mpc}$. The minimum size
of $10\textnormal{Mpc}$ was chosen to avoid entering the non-linear
physics at $<10\textnormal{Mpc}$. This configuration is shown in
Figure \ref{fig:Survey-Geometry---CTV}. In this case,
as we make our observations more sparse, the total observing time
decreases as well; we could instead choose to observe more deeply
in the same amount of time and gain volume in the redshift direction.
\begin{figure*}
\includegraphics[scale=0.3]{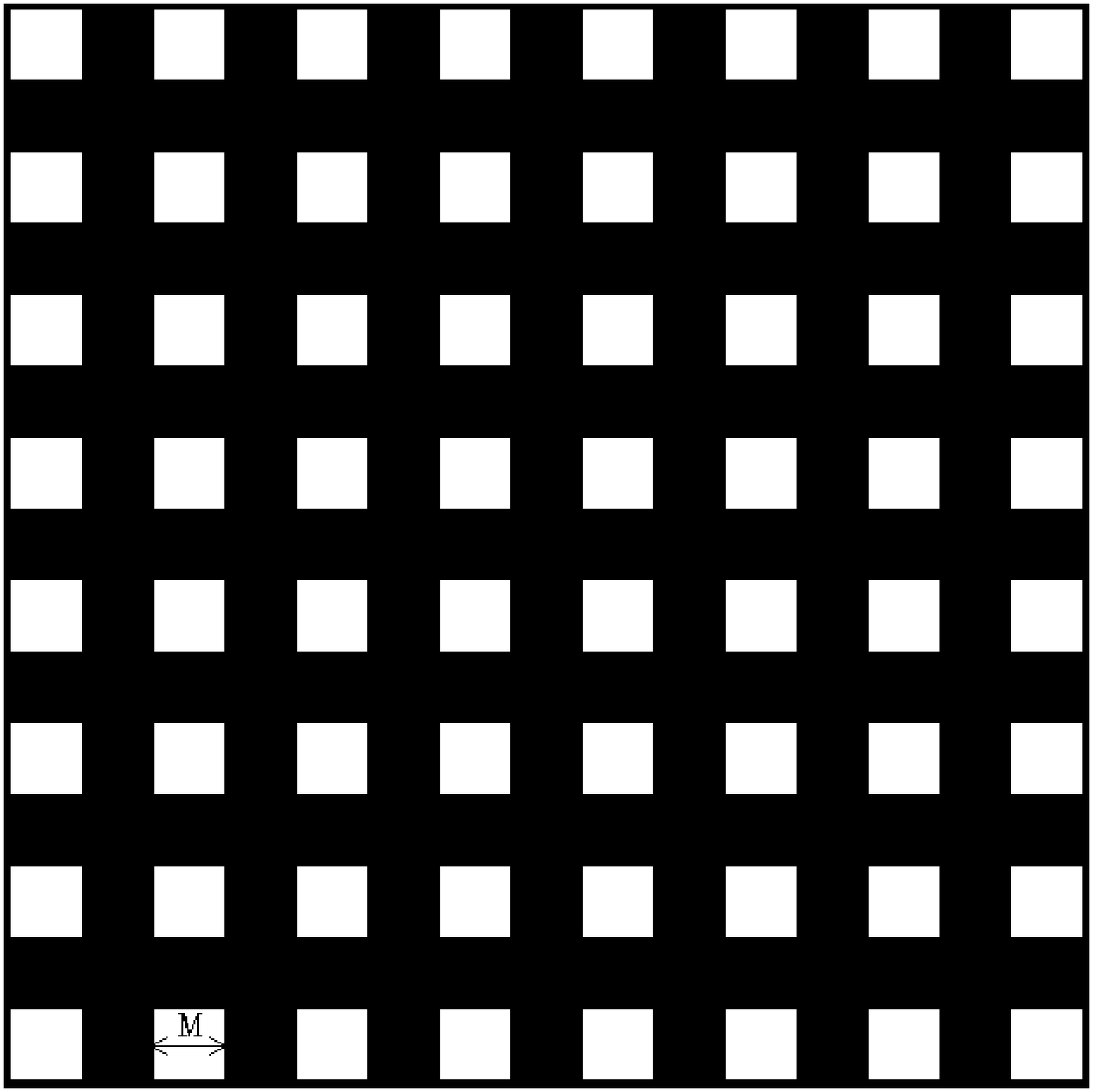}
\includegraphics[scale=0.3]{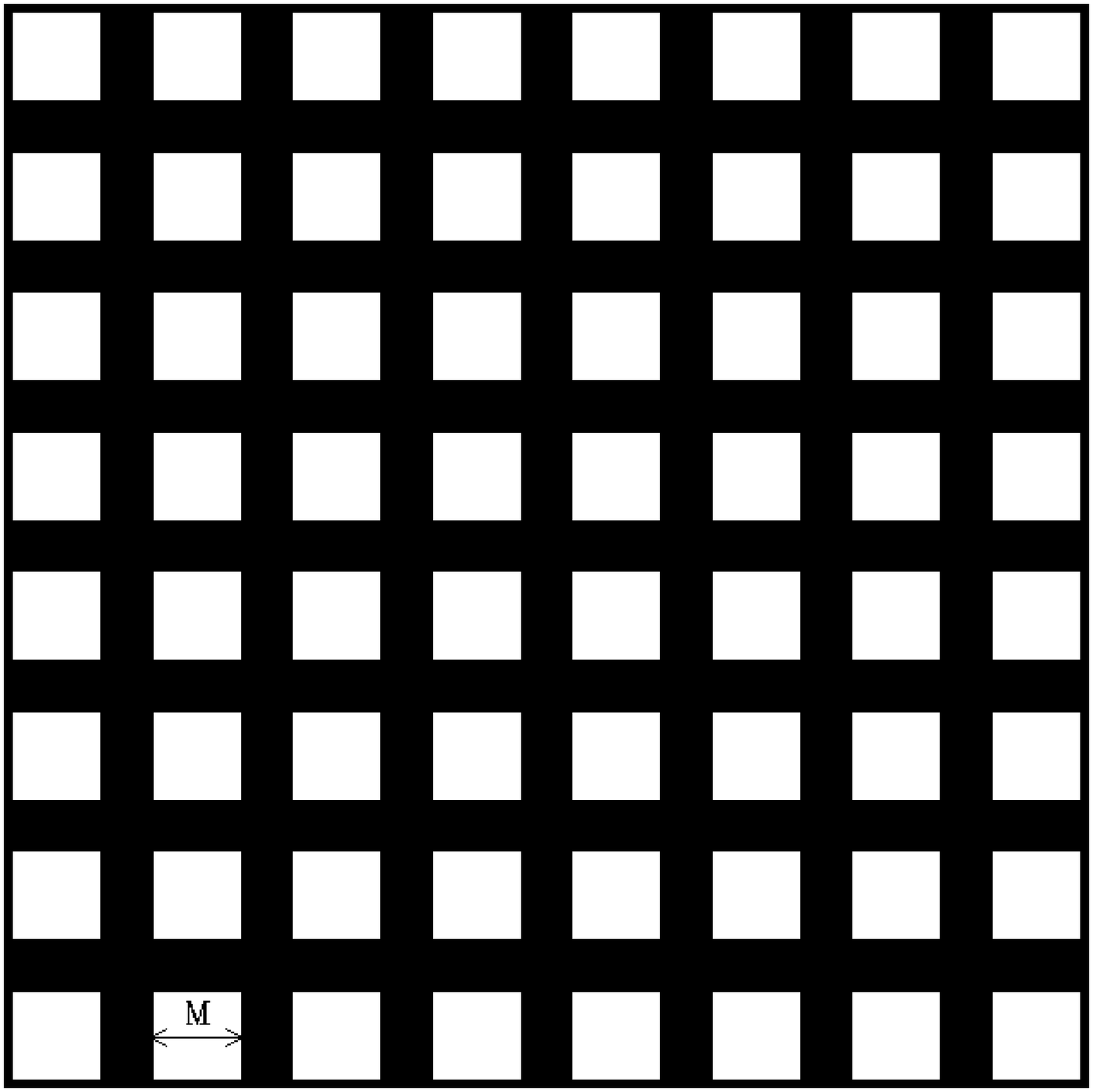}
\medskip{}
\includegraphics[scale=0.3]{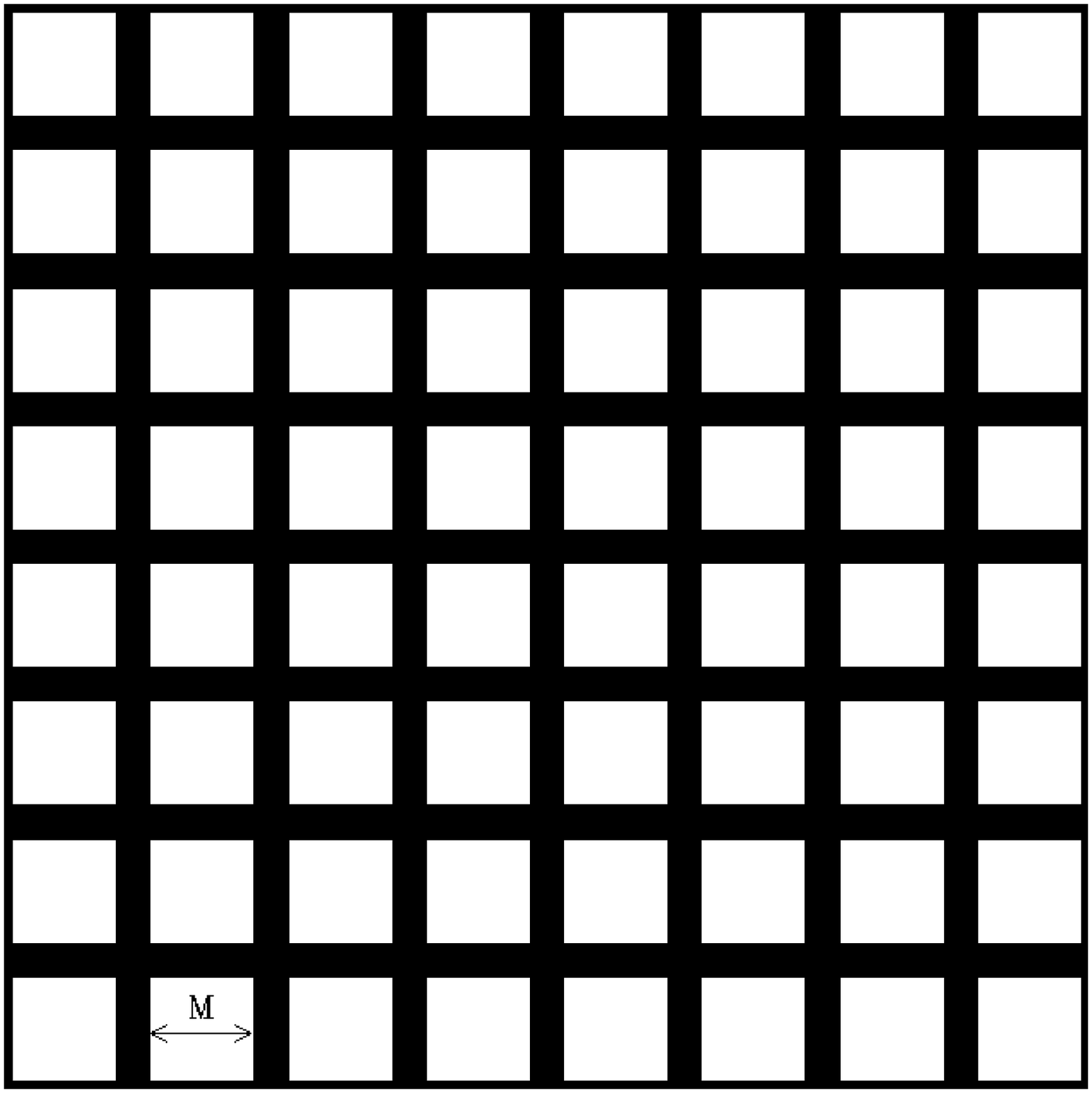}
\includegraphics[scale=0.3]{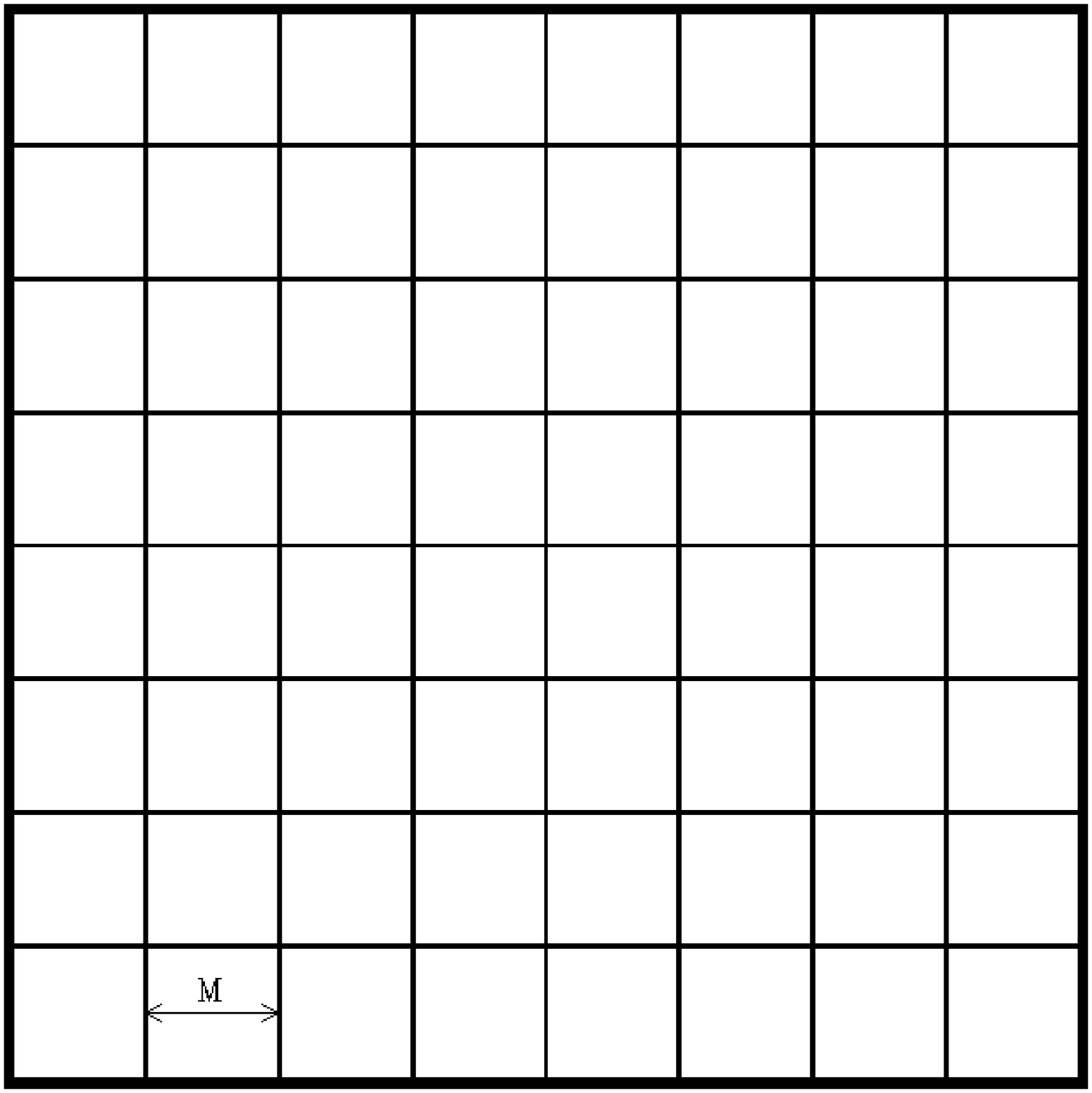}
\caption{Survey geometry for the `constant total area' scenario --- section \ref{sec:CTV}. In this setting we keep the patches at a constant position and gradually
decrease their size. Therefore, the total \textit{sampled} area (i.e., the total extent of the survey) is kept constant, while the total \textit{observed} area (and hence the survey observing time) decreases as
the patch sizes decrease. 
\label{fig:Survey-Geometry---CTV}}
\end{figure*}
\item Constant Observed Area (footprint of the survey stays constant)\\
In this setting the size of the patches are kept fixed at $60\textnormal{Mpc}$,
and the area is sparsified by placing the patches further and further
from one another. Here the total observed area is constant,
while the total sampled area increases as the patches are put further
and further. This configuration is shown in Figure \ref{fig:Survey-Geometry---COV}.
Now, the length of time for the survey remains the
same, but is spread out over a larger area of sky.\\
\begin{figure*}
\includegraphics[scale=0.4]{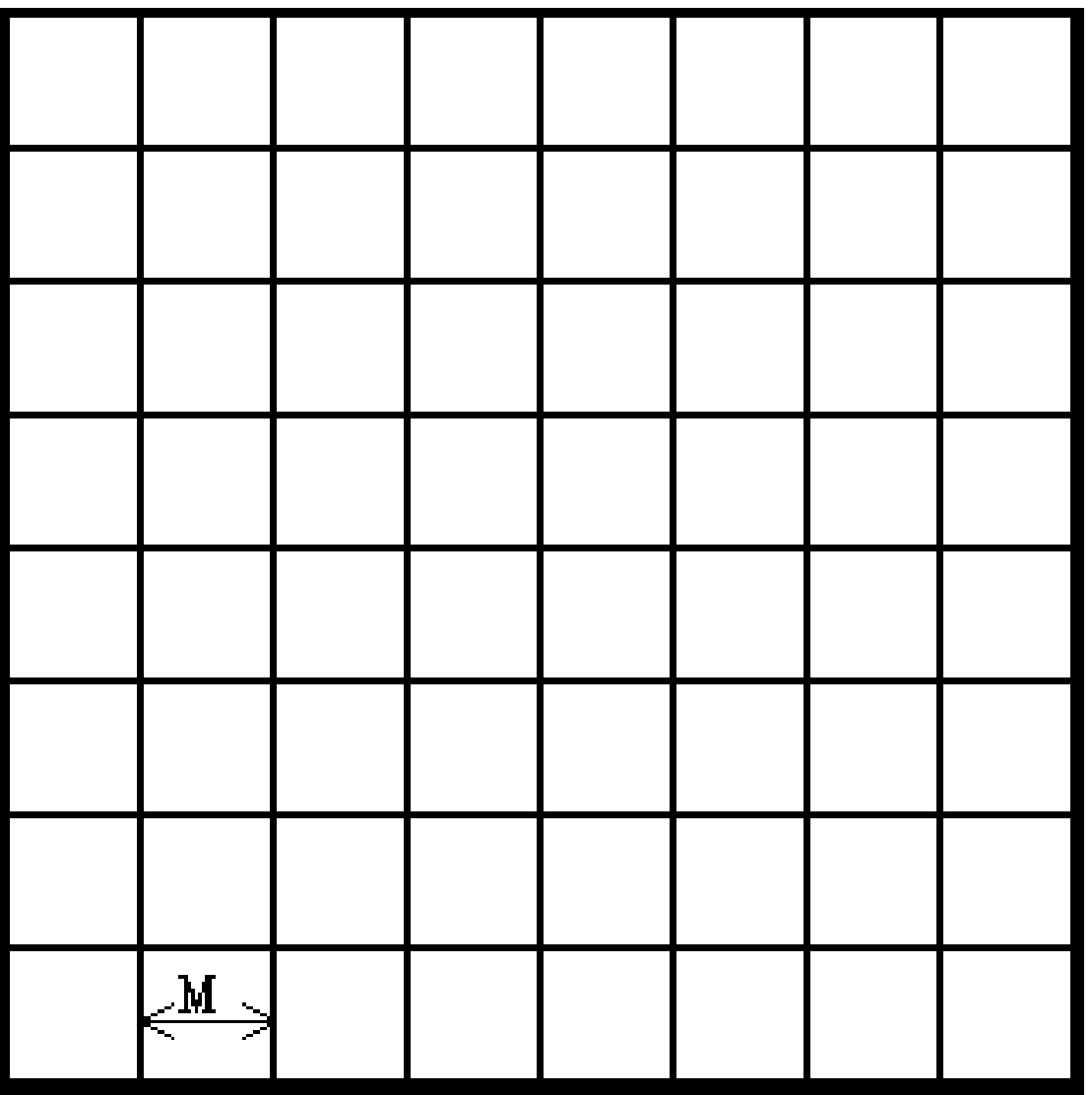}
\includegraphics[scale=0.32]{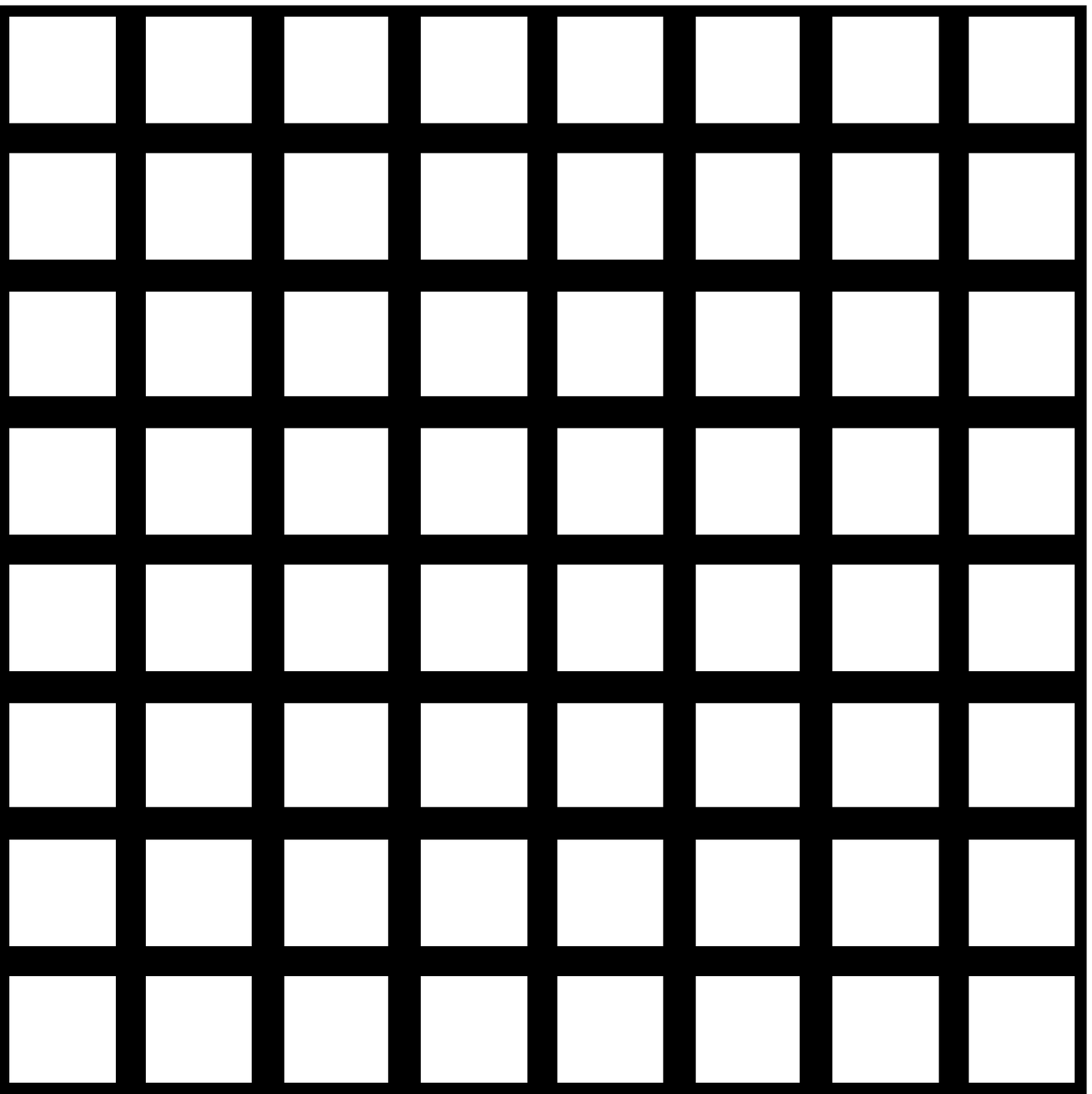}
\includegraphics[scale=0.32]{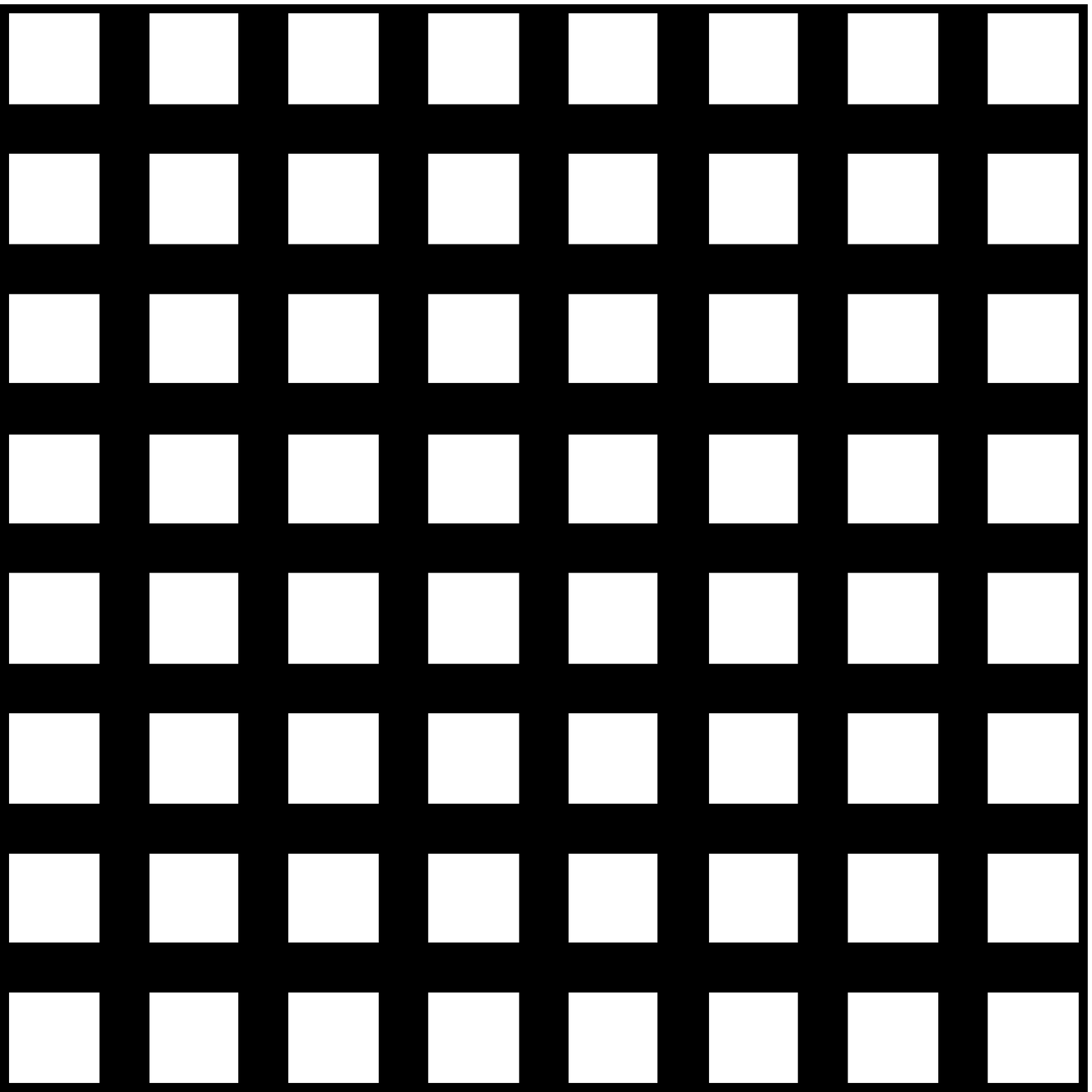}
\caption{Survey geometry for the `constant observed area' scenario --- section \ref{sec:COV}. In this setting the size of the patches are kept fixed at $60\textnormal{Mpc}$,
and the area is sparsified by placing the patches further and further
from one another. Here the total observed area (and hence the survey observing time) is constant,
while the total sampled area (i.e., the total extent of the survey) increases as the patches are put further
and further.
\label{fig:Survey-Geometry---COV}}
\end{figure*}
\end{itemize}

Note that the areas we consider here are small enough that the flat
sky approximation is valid. Also note that in all the above setting
we keep the number of bins of the galaxy power spectrum constant at
$n_{\textrm{bin}}=60$. In reality we should let the total volume
of the survey choose the binning of the power spectrum via $k_{min}=(2\pi/V)^{1/3}=dk$, and hence the number
of the bins $n_{\textrm{bin}}$. However, if $n_{bin}$ changes from
case to case it will be unfair to compare D-optimality and Entropy
as they will have different units as $n_{bin}$ changes. To have a
fair comparison between the cases we keep $n_{\textrm{bin}}$ constant.

\section{Results}

We have chosen a geometrically flat $\Lambda$CDM model with adiabatic
perturbations. We have a five-parameter model with the following values
for the parameters: $\Omega_{m}=0.214$, $\Omega_{b}=0.044$, $\Omega_{\Lambda}=0.742$,
$\tau=0.087$ and $h=0.719$, where $H_{0}=100h\textnormal{k}\textnormal{m}^{-1}\textnormal{Mp}\textnormal{c}^{-1}$.
The FoM used are 
\begin{eqnarray}
\textnormal{Entropy} & = & \left[\ln\left|\mathbf{F}\right|-\ln|\mathbf{\Pi}|-\textrm{trace}(\mathbb{I}-\mathbf{\Pi}\mathbf{F}^{-1})\right]
\nonumber\\
& & \;\times \; 0.5\,,\\
\textnormal{A-optimality} & = & \ln(\textrm{trace}(\mathbf{F}))\,,\\
\textnormal{D-optimality} & = & \ln(\left|\mathbf{F}\right|)\,,
\end{eqnarray}
where $\mathbf{\Pi}$ is the prior Fisher matrix, which we have chosen
to be that for a SDSS-LRG-like survey.
The posterior Fisher matrix is $\mathbf{F}=\mathbf{L}+\mathbf{\Pi}$,
where $\mathbf{L}$ is the likelihood Fisher matrix, which is the
current sparse survey we have designed. The utility functions above
are defined so that they need to be maximised for an optimal design.

\subsection{Constant Total Area}\label{sec:CTV}

\begin{figure*}
\includegraphics[scale=0.585]{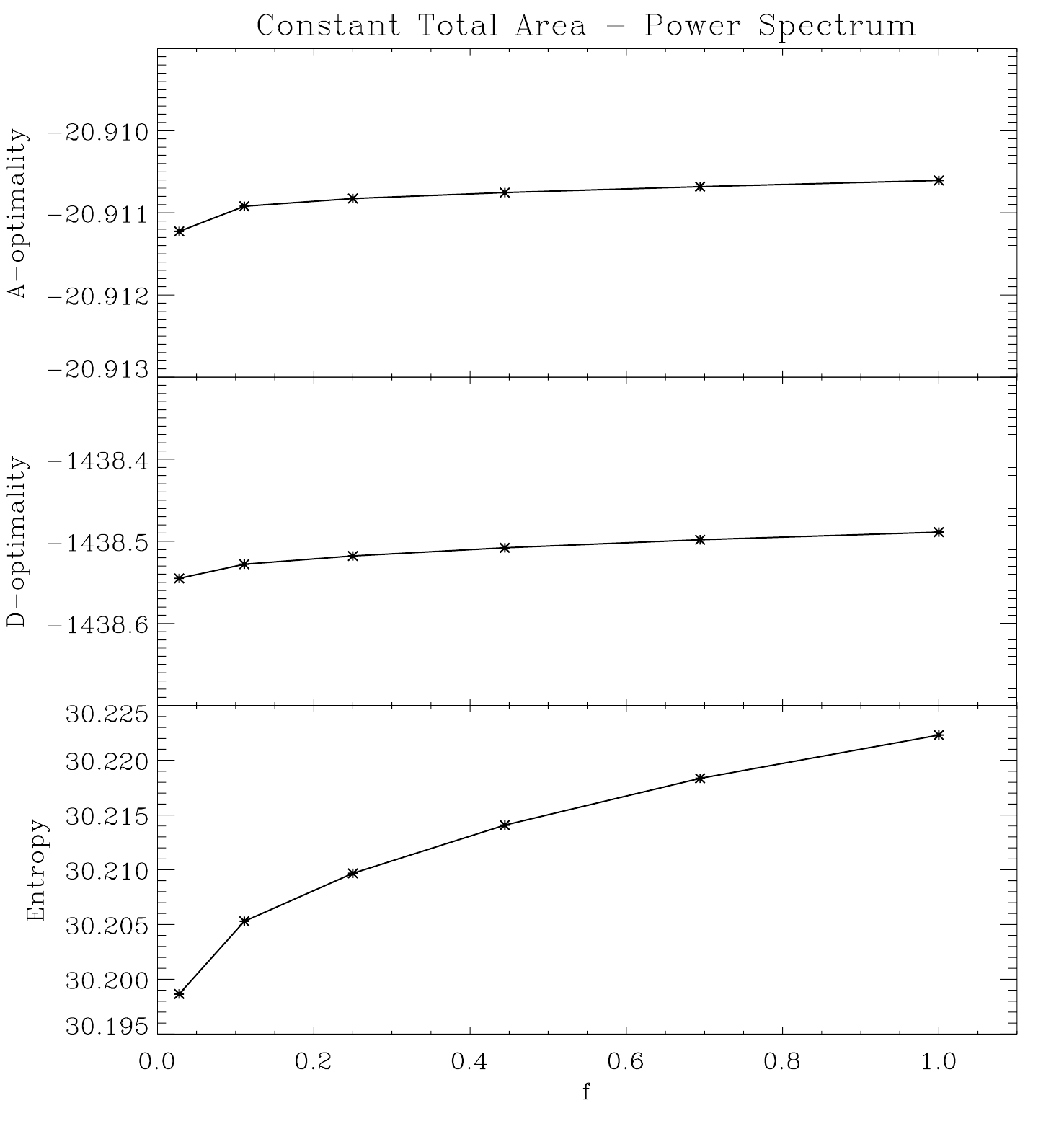}
\includegraphics[scale=0.585]{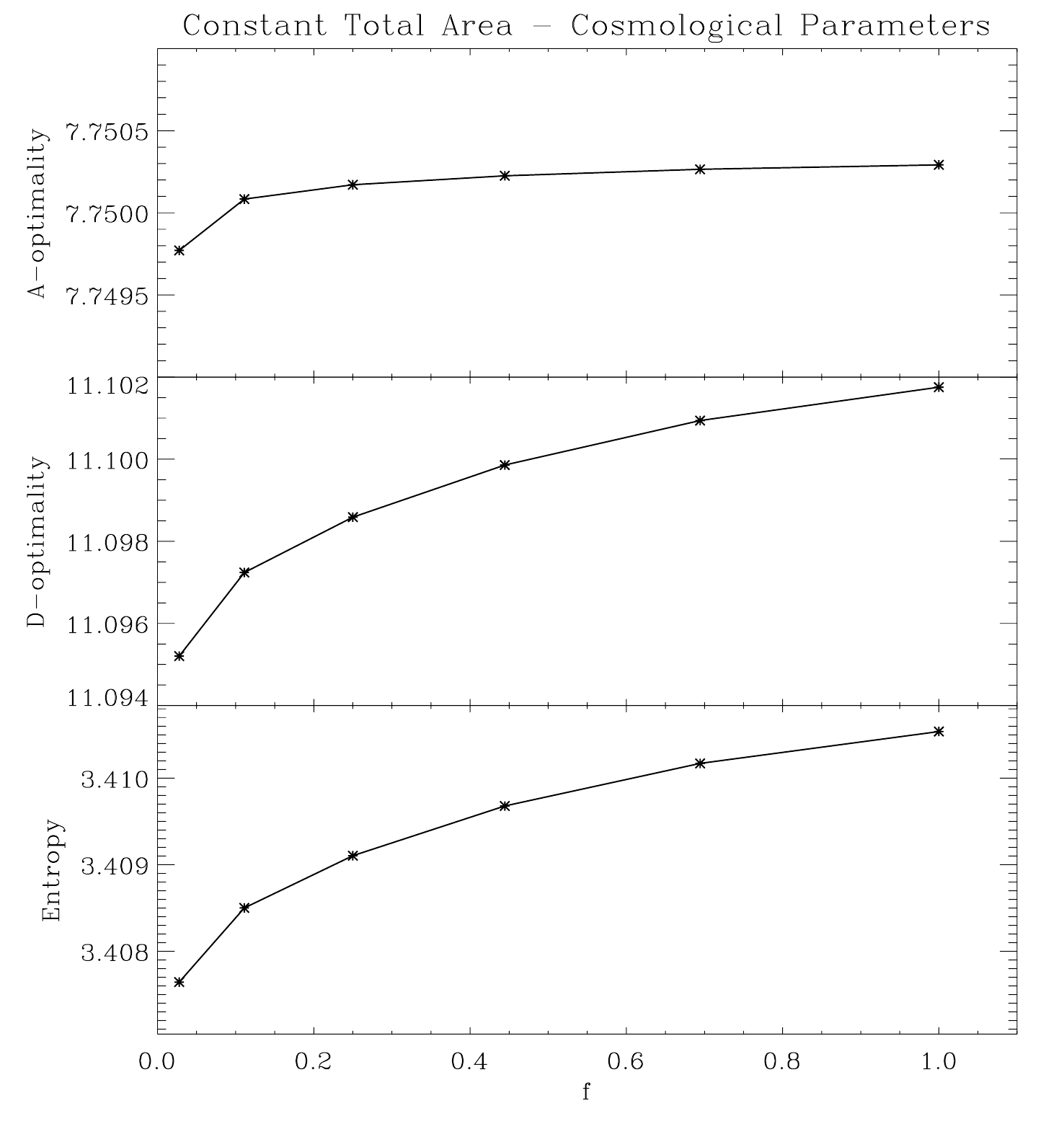}
\caption{`Constant total area' --- Figure of Merit for galaxy power spectrum bins on the left and cosmological parameters on the right. In both cases, the Entropy, A-optimality and D-optimality all increase
with $f$. This is as expected as a contiguous sampling
of the sky captures all the information and should be the best to
constrain cosmology. However, note that the increase is indeed very small. In
general we conclude that the loss in the constraining power of the survey due to sparsity
is negligible and, overall, little is gained by observing the sky more contiguously. Therefore, the sparse surveys seem like a good substitute for the contiguous surveys, with less observing time and less cost. 
\label{fig:CTV_FoM}}
\end{figure*}

\begin{figure*}
\includegraphics[scale=0.8]{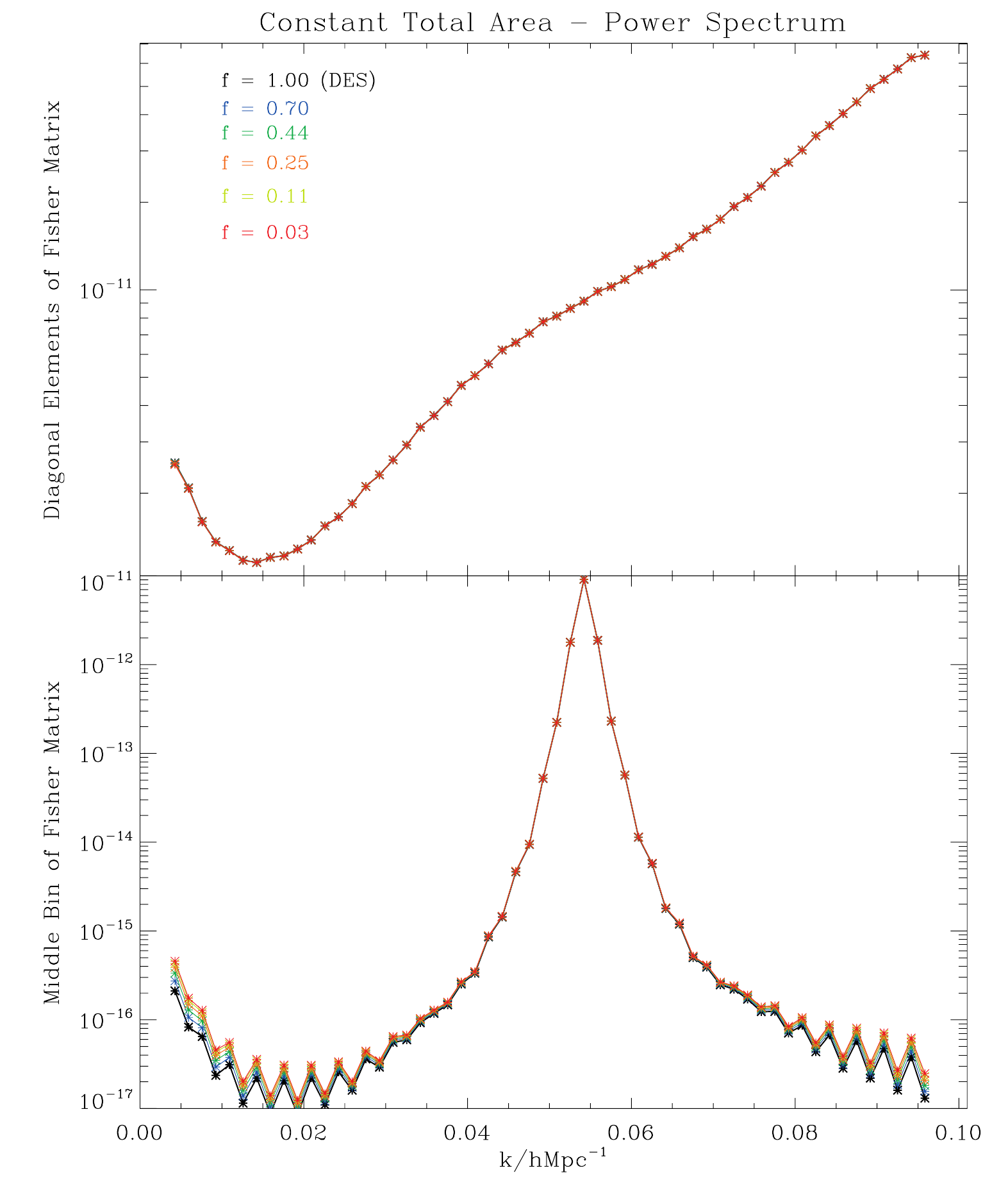}
\caption{`Constant total area' --- Top panel shows the diagonal elements of the Fisher matrix for different $f$ for the power spectrum bins. The increase in these elements (which translates into a decrease in the variance) is indeed very negligible as sparsity increases. Bottom panel shows the row of the Fisher matrix that corresponds to the middle
bin of the power spectrum. Going away from the peak in both
direction, these elements show the correlation between the different
bins and the middle one. As $f$ decreases and we get
more and more sparse, the power in the off-diagonal elements of the
Fisher matrix increases, meaning there is more aliasing between the bins. The DES survey,
as a full contiguous survey, has the least aliasing, while the sparsest
survey has the most aliasing. The uniform increase at low $k$, large scales, is due to sample variance. 
\label{fig:CTV_FM}}
\end{figure*}

Figures \ref{fig:CTV_FoM} shows the FoM for both the galaxy power
spectrum bins on the left and the cosmological parameters on the right.
In both cases, the Entropy, A-optimality and D-optimality all increase
with $f$. This is as expected as a contiguous sampling
of the sky captures all the information and should be the best to
constrain cosmology. The top panels in the Figure show A-optimality for the bins on the
left and the cosmological parameters on the right. In both cases,
$A$ increases with $f$ and reaches its maximum at
$f=1$ for DES. Note that A-optimality is a measure
of the errors of the parameters only --- it is a measure of the trace
of the Fisher matrix. Therefore, it is does not account for the correlations
between parameters. Although $A$ increases with $f$
for both the bins and the parameters, note that this increase is very
small. To see the amount of change in each of the elements of the
power spectrum Fisher matrix as $f$ increases, look at the top panel
of Figure \ref{fig:CTV_FM}. This shows the diagonal elements of the
Fisher matrix $\mathbf{F}$ for galaxy power spectrum bins for the
different $f$. The elements are all on top of each
other and indeed the gain obtained by increasing $f$
is very small. 

The middle panels of Figure \ref{fig:CTV_FoM} show D-optimality,
which again increases with $f$ for both the bins and
the parameters. Note that, D-optimality is a measure of the determinant
of the Fisher matrix and therefore takes the correlation between the
parameters into account. The correlation between the parameters is
indeed very important; one disadvantage of the sparse sampling is
the correlation it induces between the parameters due to aliasing.
To see this effect, look at the bottom panel of Figure \ref{fig:CTV_FM},
where the row of the Fisher matrix that corresponds to the middle
bin of the power spectrum is shown. Going away from the peak in both
direction, the elements show the correlation between the different
bins and the middle one. As $f$ decreases and we get
more and more sparse, the power in the off-diagonal elements of the
Fisher matrix increases, meaning there is more aliasing. The DES survey,
as a full contiguous survey, has the least aliasing, while the sparsest
survey has the most. The rise towards the small $k$ (large scales) is due to sample
variance.

Looking at the correlations and the errors in
the Fisher matrix of the spectrum one notes that the decrease in D-optimality
for sparser surveys is mostly due to the increased correlation between
the bins rather than the the increased errors; as we saw in the top
panel of this Figure the decrease in the errors are negligible. In
general we conclude that total aliasing induced by sparsity is small
and the loss in the constraining power of the survey due to this aliasing
is negligible. Hence, overall, little is gained by observing the sky more contiguously. 

The bottom panels in Figure \ref{fig:CTV_FoM} show the Entropy for the bins and
the parameters. Again, $E$ increases with $f$ and
reaches its maximum for DES. The Entropy measures the total size of
the errors of the parameters in the Fisher matrix as well as their
correlation. Hence it is a good compromise of A- and D-optimality.
It measures the total information gain of the survey relative to a
prior survey. Having an SDSS-like-survey as our prior, and taking
into account both the errors and the correlation between the parameters,
the contiguous DES survey has the largest gain compared to the sparse
surveys. However, note that this gain is again very small. 

\begin{figure}
\includegraphics[width=\columnwidth]{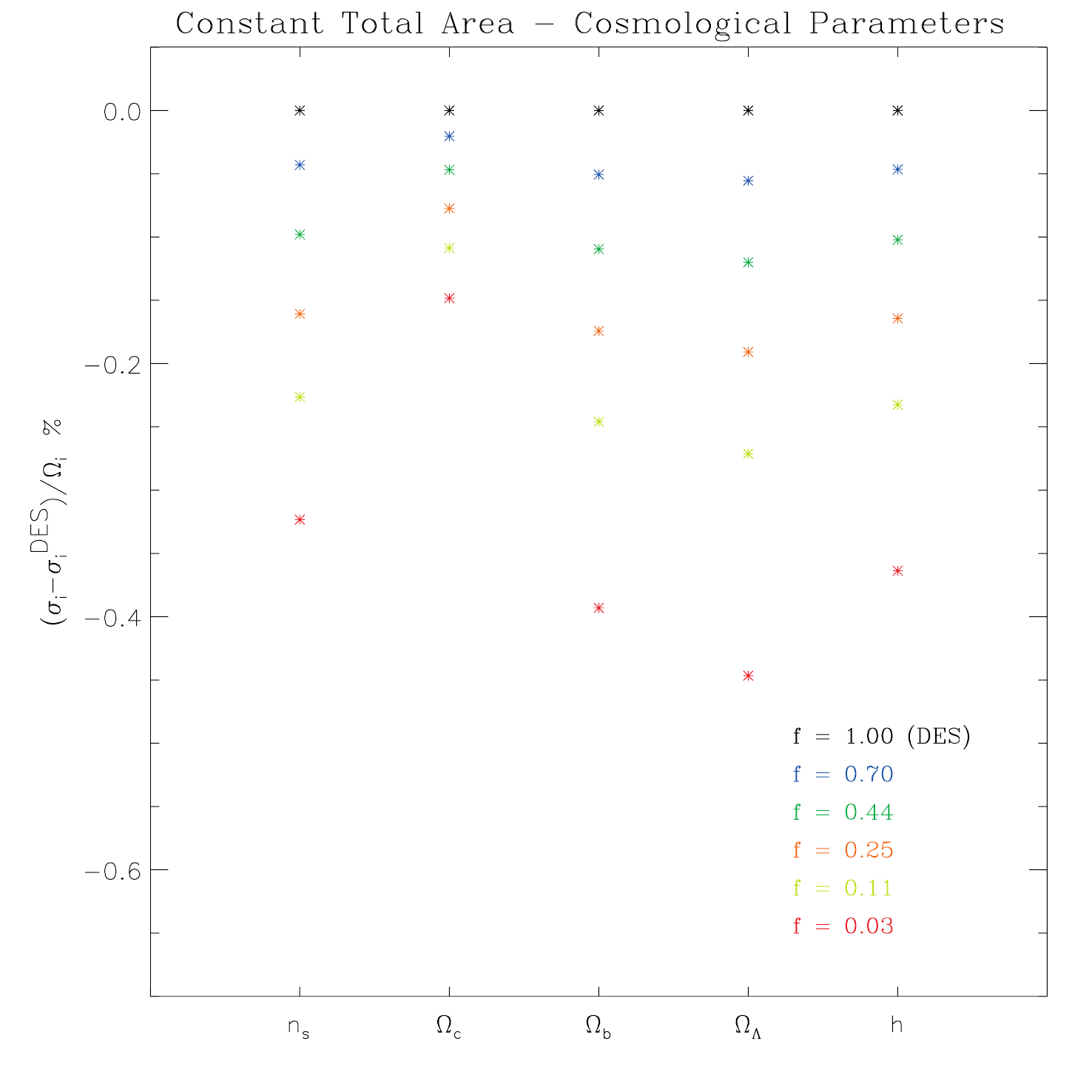}
\caption{`Constant total area'  --- Relative change in the errors of the cosmological parameters. The largest loss is about $4.5\%$ due to sparsifying the survey.
\label{fig:CTV_FMcp}}
\end{figure}

Figure \ref{fig:CTV_FMcp} shows the relative loss in the marginalised errors of
each of the cosmological parameters with respect to DES. The largest loss for a sparse observation of the sky is on the
spectral index with $\delta \Omega_\Lambda/\Omega_\Lambda\sim0.45\%$ and the smallest
is for $\Omega_{c}$ with a loss of $\delta\Omega_c/\Omega_c\sim0.15\%$. The non-marginalised errors show a qualitatively different behaviour, where $n_s$ has the largest and $\Omega_{\Lambda}$ has the smallest loss.


\subsection{Constant Observed Area}\label{sec:COV}

\begin{figure*}
\includegraphics[scale=0.585]{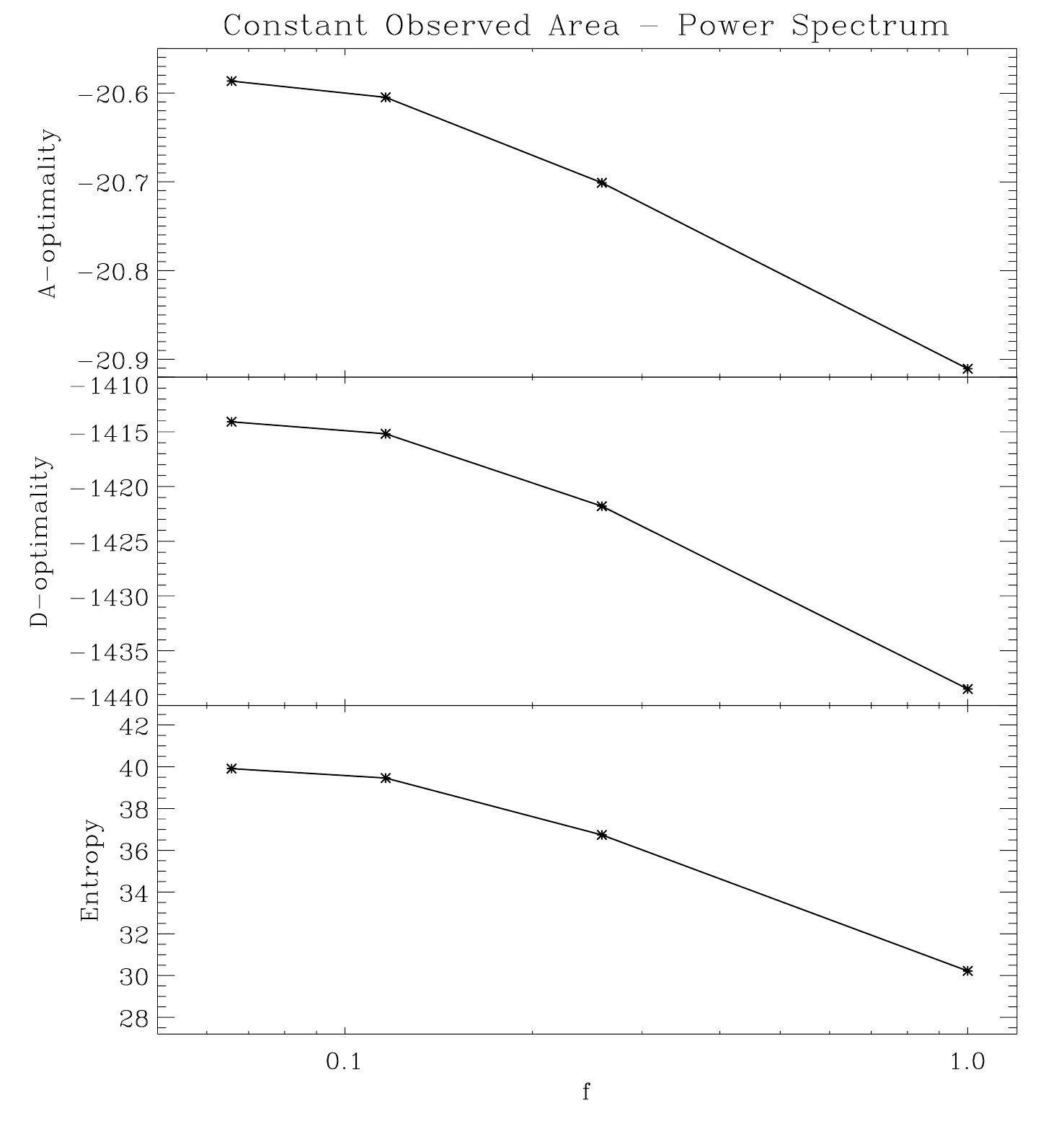}
\includegraphics[scale=0.585]{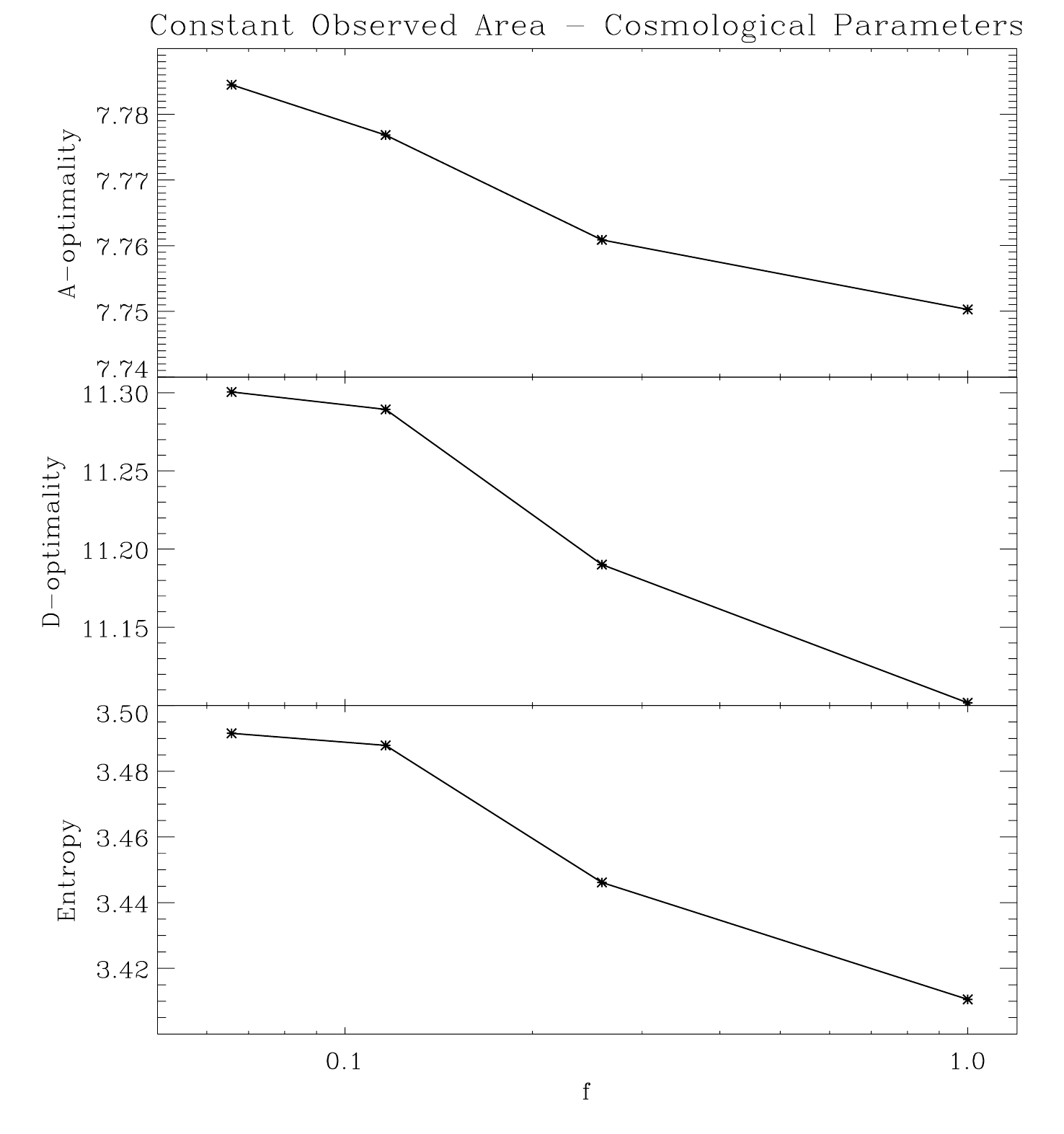}
\caption{`Constant observed area' --- Figure of Merit for galaxy power spectrum bins on the left and cosmological parameters on the right. In this case the Entropy, A-optimality
and D-optimality all decrease with $f$. And the overall
changes in all the FoM are much larger than the ones seen in the previous scenario. It seems that the increase in the total size of the
survey due to sparsity can make up for the aliasing
that the sparse design induces. Overall we gain a great deal by
spending the same amount of time on larger but sparsely sampled area. Note that the $f=0.07$ case is only for illustration purposes as it covers an area larger than the area of the sky.
\label{fig:COV_FoM}}
\end{figure*}

Figure \ref{fig:COV_FoM} shows the FoM for the power spectrum bins
and the cosmological parameters. In this case the Entropy, A-optimality
and D-optimality all decrease with $f$. And the overall
changes in all the FoM are much larger than the ones seen in the previous
scenario for both the bins and the parameters.

\begin{figure*}
\includegraphics[scale=0.8]{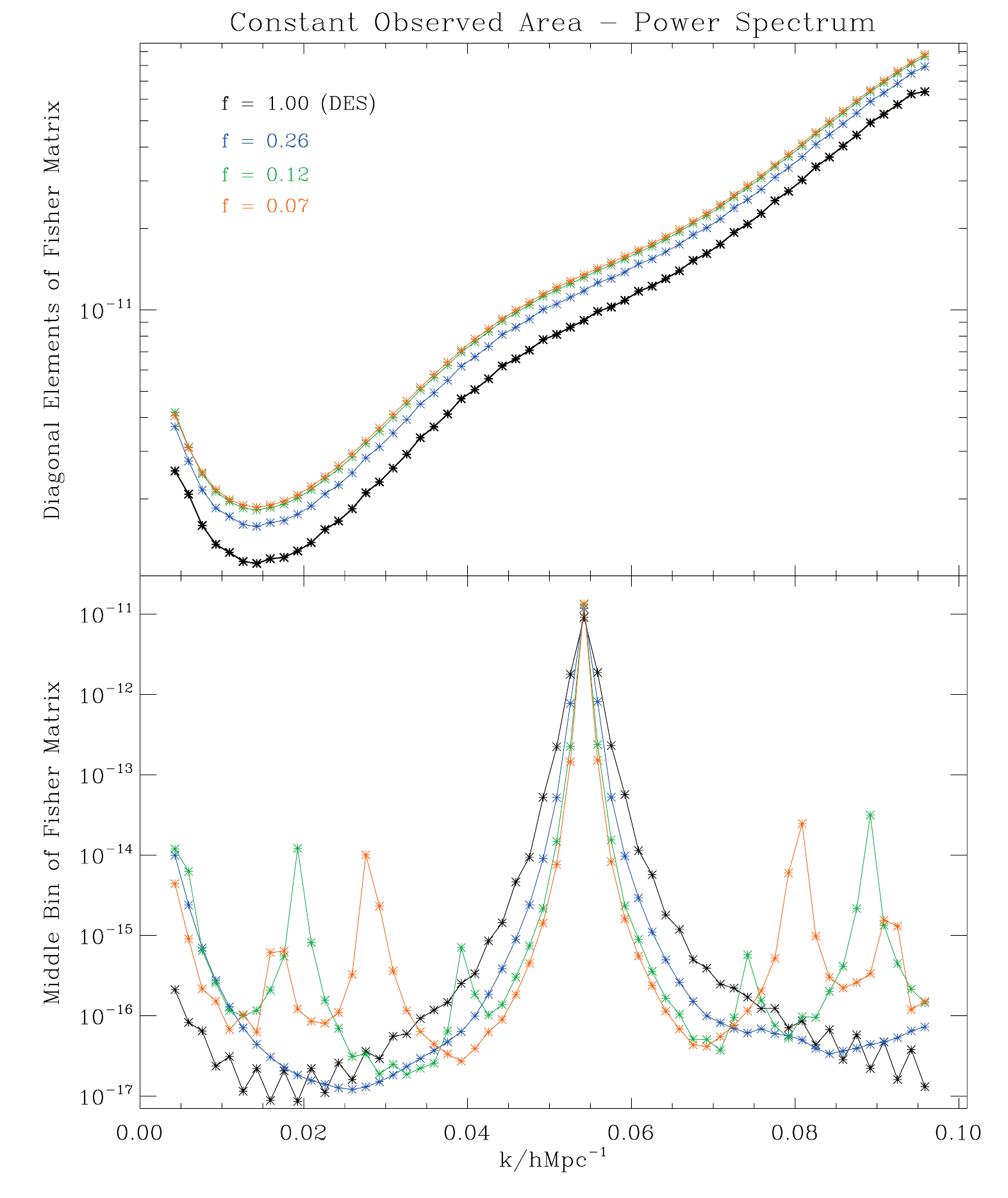}
\caption{`Constant observed area'  --- Top panel shows the diagonal elements of the Fisher matrix for different $f$ for the power spectrum bins. As we sparsify the survey these
elements increase, hence better constrain the spectrum. The bottom
panel shows the row of the Fisher matrix that corresponds
to the middle bin of the spectrum. Going away from the peak, the elements
show the correlation between the different bins and the middle one.
For DES the middle bins has a correlation with the close neighbouring
bins. But the correlation decreases as we go away from the peak. Towards
small $k$, large scales, it starts to increase again due to the sample
variance. As $f$ decreases and we get more and more
sparse, the middle bin has a sharper drop (due to the larger total
size of the survey) i.e., less correlation with neighbouring bins.
However, there is more aliasing between distant bins. Also, there
are peaks (i.e., larger correlations) at certain scales which are
related to the distances between the patches, which changes case by
case. The DES survey, as a full contiguous survey, has indeed the
least aliasing, while the sparsest survey has the most aliasing. Note that the $f=0.07$ case is only for illustration purposes as it covers an area larger than the area of the sky.
\label{fig:COV_FM}}
\end{figure*}

The top panel of Figure \ref{fig:COV_FM} shows the diagonal elements
of the Fisher matrix of the bins. As we sparsify the survey these
elements increase, and hence better constrain the spectrum. The bottom
panel in the Figure shows the row of the Fisher matrix that corresponds
to the middle bin of the spectrum. Going away from the peak, the elements
show the correlation between the different bins and the middle one.
For DES the middle bin has a correlation with the close neighbouring
bins. However, the correlation decreases as we go away from the peak. Towards
small $k$ (large scales) it starts to increase again due to sample
variance. As $f$ decreases and we get more and more
sparse, the middle bin has a sharper drop (due to the larger total
size of the survey) i.e., less correlation with neighbouring bins.
However, there is more aliasing between distant bins. Also, there
are peaks (i.e., larger correlations) at certain scales which are
related to the distances between the patches, which changes case by
case. The DES survey, as a full contiguous survey, has indeed the
least aliasing, while the sparsest survey has the most. 

Note that in this case the sparsity is obtained by placing the observed
patches further and further away from each other. As the sparsity
increases as the patches are placed further, the total size of the
survey is greatly increased, which seems to make up for the aliasing
that the sparse design has induced. Overall we gain a great deal by
spending the same amount of time on larger but sparsely sampled area. 

\begin{figure}
\includegraphics[width=\columnwidth]{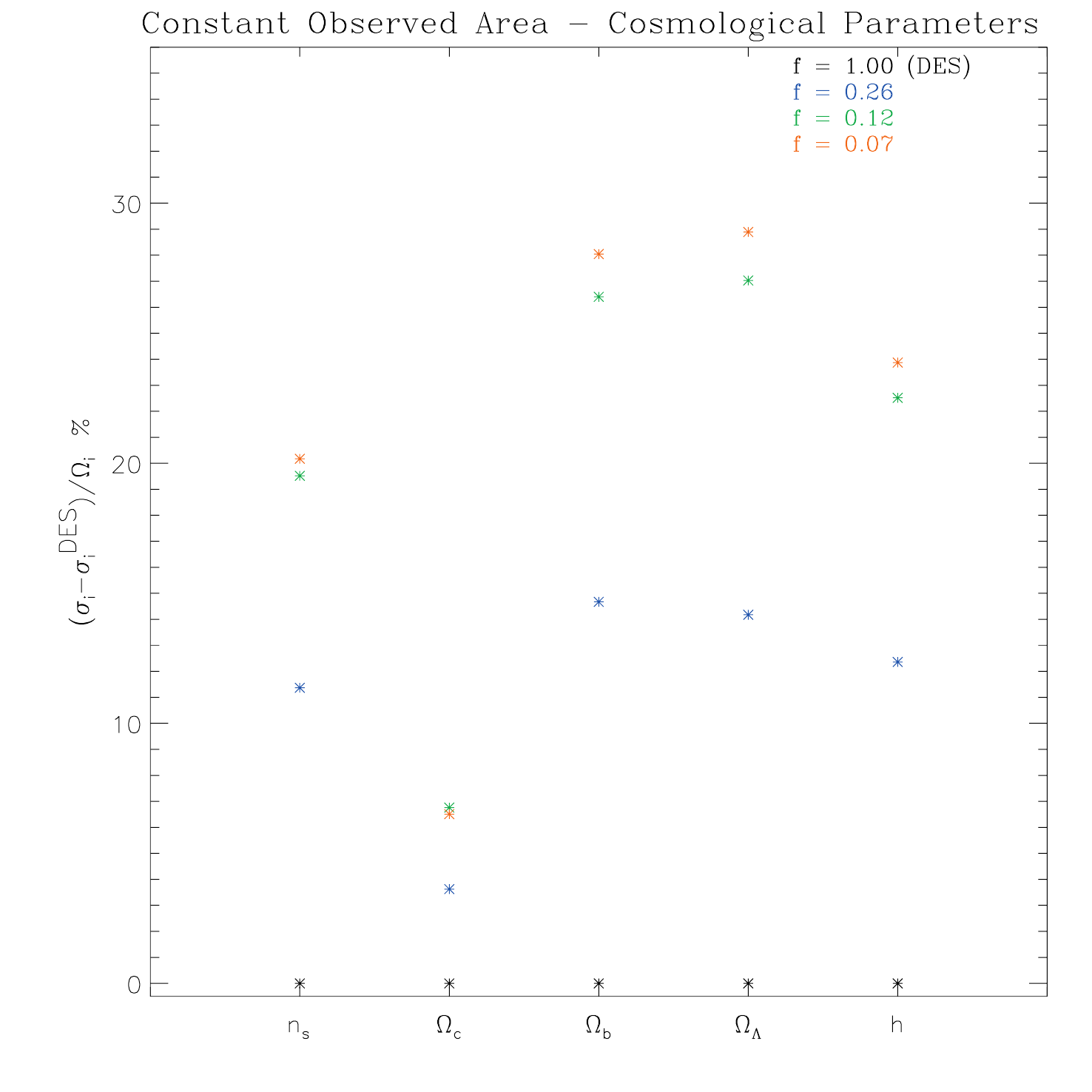}
\caption{`Constant observed area' --- Relative change in the errors of the cosmological parameters. The largest gain is about $27\%$ by sparsifying the survey. Note that the $f=0.07$ case is only for illustration purposes as it covers an area larger than the area of the sky.
\label{fig:COV_FMcp}}
\end{figure}

Figure \ref{fig:COV_FMcp} shows the relative gain in the marginalised errors of
each of the cosmological parameters with respect to DES. The largest gain for a sparse observation of the
sky is on $\Omega_{\Lambda}$ with $\delta\Omega_{\Lambda}/\Omega_{\Lambda}\sim27\%$
and the smallest is for $\Omega_c$ with a gain of $\delta\Omega_{c}/\Omega_{c}\sim7\%$.
Again, a qualitatively different scenario is seen for the non-marginalised errors; $\Omega_{b}$ has
the largest gain due to sparsity, and $h$ has the smallest. 


\section{Conclusion}

In this work we have investigated the advantages and disadvantages
of sparsely sampling the sky as opposed to a contiguous observation.
By making use of Bayesian Experimental Design, we have defined our
Figure of Merit as different functions of the Fisher matrix. These
FoM capture different aspects of the parameters of interest such as
their overall variance, the correlation between them or a measure of
both as in Entropy. By optimising these functions we investigate
an optimal survey design for estimating the galaxy power spectrum and a set of
cosmological parameters. We have compared a series of sparse designs
to a contiguous design of DES. We split the area of the DES survey into
small square patches and sparsify the survey in two ways:
\begin{enumerate}
\item by shrinking the size of the patches while they
are kept at a constant position. In this case the total sampled area
of the survey is constant while the observed area (and the survey
observing time) shrinks. This means the total information gained from the survey reduces in each case. In this scenario all the three FoM (A-optimality,
D-optimality and Entropy) increase with $f$, both
for the power spectrum bins and the cosmological parameters. This
is expected as a contiguous sampling should capture all the information
and constrain cosmology the best. However, we note that this increase with decreasing
sparsity is very small for both the bins and the cosmological parameters.
Looking at the variance and the covariance of the parameters, we note
that the slight degrading of the surveys due to sparsity is mostly because
of the increased correlation between the bins --- aliasing --- rather than the the increased
errors. In general we conclude that total aliasing induced by sparsity
is small and the loss in the constraining power of the survey because of it is negligible. Hence, overall, little
is gained by observing the sky more contiguously. Indeed the largest loss in terms of the errors of the cosmological parameters is of the order of $\sim4.5\%$ in the sparsest case.

\item by keeping the size of the patches constant, but
placing them further and further from one another. In this scenario
the observed area (and observing time) is kept constant, while sparsifying means larger
and larger total sampled area. This means the total information gained from the survey in each case is the same. Therefore, there are the two competing factors; one is the increase in the total sampled area as the survey is sparsified and the other is aliasing induced due to the larger and larger sparse mask on the sky. 

In this case all FoM decrease with
$f$, and the change in the FoM is much larger
than the ones seen in the previous scenario. As we sparsify the survey
the decrease in errors makes up for the increased aliasing induced and hence
cause a general improvement in constraining power of the survey.
Overall we gain a great deal by spending the same amount of time on
larger but sparsely sampled area. Indeed we gain as much as $\sim27\%$ on the sparsest survey, which is a significant improvement. 
\end{enumerate}
We conclude that sparse sampling could be a good substitute
for the contiguous observations and indeed the way forward for future
surveys. At least for small areas of the sky, such as that of DES,
sparse sampling of the sky can have less cost and less observing time,
while obtaining the same amount of constraints on the cosmological
parameters. On the other hand we can spend the same amount of time but sparsely observe a larger area of the sky. This greatly
improves the constraining power of the survey. 

In this work we have chosen square observation patches, which may
be the worst shape in terms of the correlation they induce. Yet another
constraint in this design is the fixed and determined positions of
the patches which cause a loss of information at certain scales. The
advantage of this approach has been its analytical formalism, which
has made it possible to understand the important factors in the sparse
sampling. For future work we will investigate an optimal shape foe the patches and have a numerical approach where
these patches are randomly distributed on the sky. This causes an even
loss of information on all scales and is expected to improve results
greatly. 

\bibliographystyle{mn2e}
\bibliography{biblio}

\begin{thebibliography}{}

\bibitem[\protect\citeauthoryear{Adelman-McCarthy et~al.,}{Adelman-McCarthy
  et~al.}{2008}]{sdss}
Adelman-McCarthy J.~K.,  et~al., 2008, Astrophys. J. Suppl., 175, 297

\bibitem[\protect\citeauthoryear{{Blake}, {Parkinson}, {Bassett}, {Glazebrook},
  {Kunz} \& {Nichol}}{{Blake} et~al.}{2006}]{Blakeetal06}
{Blake} C.,  {Parkinson} D.,  {Bassett} B.,  {Glazebrook} K.,  {Kunz} M.,
  {Nichol} R.~C.,  2006, MNRAS, 365, 255

\bibitem[\protect\citeauthoryear{{Bond}, {Jaffe} \& {Knox}}{{Bond}
  et~al.}{1998}]{bjk}
{Bond} J.~R.,  {Jaffe} A.~H.,    {Knox} L.,  1998, PRD, 57, 2117

\bibitem[\protect\citeauthoryear{Croom et~al.,}{Croom  et~al.}{2004}]{2dF}
Croom S.~M.,  et~al., 2004, Mon. Not. Roy. Astron. Soc., 349, 1397

\bibitem[\protect\citeauthoryear{{Kaiser}}{{Kaiser}}{1984}]{kaiser1984_bias}
{Kaiser} N.,  1984, APJL, 284, L9

\bibitem[\protect\citeauthoryear{{Kaiser}}{{Kaiser}}{1986}]{Kaiser-Sparse}
{Kaiser} N.,  1986, MNRAS, 219, 785

\bibitem[\protect\citeauthoryear{{Laureijs}}{{Laureijs}}{2009}]{Euclid}
{Laureijs} R.,  2009, ArXiv e-prints

\bibitem[\protect\citeauthoryear{{Liddle}, {Mukherjee} \& {Parkinson}}{{Liddle}
  et~al.}{2006}]{Liddleetal06}
{Liddle} A.,  {Mukherjee} P.,    {Parkinson} D.,  2006, Astronomy and
  Geophysics, 47, 040000

\bibitem[\protect\citeauthoryear{{Parkinson}, {Blake}, {Kunz}, {Bassett},
  {Nichol} \& {Glazebrook}}{{Parkinson} et~al.}{2007}]{PBKBNG-BayesExp}
{Parkinson} D.,  {Blake} C.,  {Kunz} M.,  {Bassett} B.~A.,  {Nichol} R.~C.,
  {Glazebrook} K.,  2007, MNRAS, 377, 185

\bibitem[\protect\citeauthoryear{{Peebles}}{{Peebles}}{1973}]{Peebles1973}
{Peebles} P.,  1973, APJ, 185, 413

\bibitem[\protect\citeauthoryear{{Tegmark}}{{Tegmark}}{1997}]{tegmark1997}
{Tegmark} M.,  1997, Physical Review Letters, 79, 3806

\bibitem[\protect\citeauthoryear{{The Dark Energy Survey Collaboration}}{{The
  Dark Energy Survey Collaboration}}{2005}]{DES}
{The Dark Energy Survey Collaboration} 2005, ArXiv Astrophysics e-prints

\bibitem[\protect\citeauthoryear{{Trotta}}{{Trotta}}{2007a}]{Trotta07}
{Trotta} R.,  2007a, MNRAS, 378, 72

\bibitem[\protect\citeauthoryear{{Trotta}}{{Trotta}}{2007b}]{Trotta07-BayesFactor}
{Trotta} R.,  2007b, MNRAS, 378, 819

\bibitem[\protect\citeauthoryear{{Trotta}}{{Trotta}}{2007c}]{T-BayesExp}
{Trotta} R.,  2007c, MNRAS, 378, 819

\end{thebibliography}

\label{lastpage}
\end{document}